\documentclass[
reprint,
superscriptaddress,
nofootinbib,
nobibnotes,
amsmath,
amssymb,
aps,
prd,
floatfix,
]{revtex4-2}

\usepackage{graphicx}
\usepackage{subfigure}
\usepackage{dcolumn}
\usepackage{bm}
\usepackage{xspace}
\usepackage{hyperref}
\hypersetup{
    colorlinks,
    linkcolor={red!50!black},
    citecolor={blue!90!black},
    urlcolor={blue!80!black}
}

\usepackage[utf8]{inputenc} 
\usepackage[T1]{fontenc}    
\usepackage{hyperref}       
\usepackage{booktabs}       
\usepackage{amsfonts}       
\usepackage{nicefrac}       
\usepackage{microtype}      
\usepackage[table]{xcolor}  
\makeatletter
\AtBeginDocument{\appdef\tabular@hook{\col@sep\tabcolsep}}
\makeatother
\usepackage{amsmath}
\usepackage{graphicx}
\usepackage{subfigure}
\usepackage{bm}
\usepackage{booktabs}
\usepackage{comment}
\usepackage[capitalise]{cleveref}

\usepackage[printonlyused]{acronym}

\usepackage{tikz}
\usetikzlibrary{arrows.meta,positioning,calc,fit}

\newcommand{\bilby}{\texttt{bilby}\xspace}

\newcommand{\dynesty}{\texttt{dynesty}\xspace}

\newcommand{\BayesStar}{\texttt{BayesStar}\xspace}
\newcommand{\aspire}{\ac{ASPIRE}\xspace}
\newcommand{\pocomc}{\texttt{pocomc}\xspace}
\newcommand{\minipcn}{\texttt{minipcn}\xspace}
\newcommand{\simplepe}{\texttt{simple-pe}\xspace}
\newcommand{\scipy}{\texttt{SciPy}\xspace}

\newacro{CP}{Conformal Prediction}
\newacro{MCP}{Mondrian Conformal Prediction}
\newacro{CBC}{Compact Binary Coalescence}
\newacro{FAR}{False Alarm Rate}
\newacro{IFAR}{Inverse \ac{FAR}}
\newacro{MDC}{Mock Data Challenge}
\newacro{ROC}{Receiver Operator Curve}
\newacro{AUC}{Area Under Curve}
\newacro{ML}{Machine Learning}
\newacro{MLP}{Multi-layer Percepton}
\newacro{LR}{Logistic Regression}
\newacro{NN}{Neural Network}
\newacro{CNN}{Convolutional Neural Network}
\newacro{LVK}{LIGO-Virgo-KAGRA}
\newacro{GWTC}{Gravitational-Wave Transient Catalog}
\newacro{LHO}{LIGO Hanford Observatory}
\newacro{LLO}{LIGO Livingston Observatory}
\newacro{BBH}{Binary Black Hole}
\newacro{BNS}{Binary Neutron Star}
\newacro{MCMC}{Markov Chain Monte Carlo}
\newacro{SMC}{Sequential Monte Carlo}
\newacro{ASPIRE}{Accelerated Sequential Posterior Inference via Reuse}
\newacro{JSD}{Jensen--Shannon Divergence}
\newacro{PP}[P--P]{Probability--Probability}
\newacro{SNR}{Signal-to-Noise Ratio}
\newacro{MAP}{Maximum A Posteriori}
\newacro{DE}{Differential Evolution}
\newacro{FIM}{Fisher Information Matrix}
\newacro{DALI}{Derivative Approximation for LIkelihoods}
\newacro{PSD}{Power Spectral Density}
\newacro{LISA}{Laser Interferometer Space Antenna}

\newcommand{\parameters}{\ensuremath{\theta}\xspace}
\newcommand{\parametersstar}{\ensuremath{\hat{\parameters}}\xspace}
\newcommand{\parameterstrue}{\ensuremath{\parameters_{\rm true}}\xspace}
\newcommand{\parametersinj}{\ensuremath{\parameters_{\rm inj}}\xspace}
\newcommand{\data}{\ensuremath{d}\xspace}
\newcommand{\signal}{\ensuremath{h}}
\newcommand{\noise}{\ensuremath{n}\xspace}
\newcommand{\likelihood}{\ensuremath{\mathcal{L}}\xspace}
\newcommand{\prior}{\ensuremath{\pi}\xspace}
\newcommand{\evidence}{\ensuremath{\mathcal{Z}}\xspace}
\newcommand{\posterior}{\ensuremath{p}\xspace}
\newcommand{\model}{\ensuremath{\mathcal{H}}\xspace}
\newcommand{\potential}{\ensuremath{\Phi}\xspace}
\newcommand{\unnorm}{\ensuremath{\tilde{\posterior}}\xspace}
\newcommand{\N}{\ensuremath{\mathcal{N}}\xspace}

\newcommand{\mat}[1]{\ensuremath{\mathbf{#1}}}
\newcommand{\transpose}{\ensuremath{\mathsf{T}}}
\newcommand{\hessian}{\ensuremath{\mat{H}}\xspace}

\newcommand{\inner}[2]{\ensuremath{\left\langle #1 \mathrel{\big|} #2 \right\rangle}}
\newcommand{\noiseavg}[2][]{\ensuremath{\textrm{E}_{#1}\left[#2 \right]}\xspace}
\newcommand{\fisher}{\ensuremath{\mat{\Gamma}}\xspace}

\newcommand{\nlikelihood}{\ensuremath{N_{\rm l}}\xspace}
\newcommand{\nrestarts}{\ensuremath{n_{\rm r}}\xspace}
\newcommand{\nparameters}{\ensuremath{k}\xspace}

\begin{document}

\title{Fisher and Bayes: accelerated gravitational-wave inference with model-informed seeding}

\author{Gregory Ashton}
\email{G.Ashton@soton.ac.uk}
\affiliation{Mathematical Sciences, University of Southampton, Southampton SO17 1BJ, United Kingdom}
\affiliation{Department of Physics, Royal Holloway, University of London, Egham, Surrey TW20 0EX, United Kingdom}
\author{Michael J. Williams}

\affiliation{Max Planck Institute for Gravitational Physics (Albert Einstein Institute), Am M\"uhlenberg 1, 14476 Potsdam, Germany}
\affiliation{Institute of Cosmology and Gravitation, University of Portsmouth, Portsmouth, PO1 3FX, United Kingdom}

\date{\today}

\begin{abstract}
Bayesian inference of transient gravitational-wave signals typically relies on stochastic samplers initialised from the prior.
For signals with a high signal-to-noise ratio, which are increasingly common as detector sensitivity improves, this makes exploration of the parameter space inefficient.
We instead seed sampling from a Gaussian mixture model, with modes located by differential evolution and covariances set by the Fisher information matrix.
We use the mixture in four ways: as a direct posterior approximation truncated to the prior support; as a proposal for rejection sampling; as the initial distribution for sequential Monte Carlo; and as an effective prior for nested sampling via posterior repartitioning.
The direct approximation is near-instantaneous and provides a useful first look, but it is not statistically consistent with reference posteriors, even at high signal-to-noise ratio.
The other three methods correct for the approximation, achieving speed-ups of up to a factor of $\sim 4$ over nested sampling on gravitational-wave examples, with a trade-off between speed and accuracy.
Simulation studies quantify this trade-off, and a large-scale probability--probability test confirms that Laplace-seeded sequential Monte Carlo is unbiased, if properly configured.
However, we find that seeding of a poor quality, such as missed modes, can diminish efficiency and accuracy.
Overall, we conclude that model-informed seeding can provide modest acceleration and a rapid first look at the posterior distribution for gravitational-wave signals.
\end{abstract}

\maketitle

\section{Introduction}
\label{sec:introduction}
Gravitational waves observed by ground-based interferometers such as LIGO\footnote{The two LIGO detectors: the \ac{LHO} and the \ac{LLO}.} \citep{LIGOScientific:2014pky}, Virgo \citep{VIRGO:2014yos}, and KAGRA \citep{KAGRA:2020tym} carry rich information about the astrophysical sources that generate them.
To date, the \ac{LVK} observatories have discovered several hundred \ac{CBC} signals from systems containing black holes and neutron stars, published in the \ac{GWTC} \citep{LIGOScientific:2026sit, LIGOScientific:2026wfs}.

Since the early days of the field, the process of searching for signals and measuring their significance has been separated from the process of inferring the properties of their sources \citep{Finn:1992wt}.
This latter task, \emph{parameter estimation}, involves approximating the multi-dimensional posterior distribution of the parameters $\parameters$ given the data $\data$ and a model $\model$, which by Bayes' theorem is
\begin{align}
    \posterior(\parameters|\data, \model) = \frac{\likelihood(\data|\parameters, \model) \prior(\parameters|\model)}{\evidence(\data|\model)}\,,
    \label{eq:posterior}
\end{align}
where $\likelihood$ is the likelihood, $\prior$ is the prior and $\evidence$ is the evidence:
\begin{align}
    \evidence(\data|\model) = \int d\parameters \likelihood(\data|\parameters, \model) \prior(\parameters|\model)\,.
    \label{eq:evidence}
\end{align}

State-of-the-art inference methods use stochastic sampling to approximate the posterior distribution and evidence.
Specifically, nested sampling \citep{2004AIPC..735..395S} and \ac{MCMC} \citep{Metropolis:1953am, Hastings:1970aa} have been widely deployed \citep[see, e.g.][]{Veitch:2014wba, Ashton:2018jfp,Biwer:2018osg,Roulet:2022kot}.
As current detectors improve in sensitivity and next-generation detectors are built, both the number of observed signals and their \ac{SNR} will increase.
This drives the development of faster algorithms, to handle the growing data volume, to reduce the time to produce results (enabling iterative studies), and to make computationally expensive models tractable.
Broadly, inference can be accelerated by faster likelihood evaluation \citep{Canizares:2014fya, Cornish:2021lje, Zackay:2018qdy}, better sampling efficiency \citep{Roulet:2022kot, Thrane:2018qnx}, parallelisation \citep{Smith:2019ucc, Prathaban:2025qgg}, or a combination of these.

Before stochastic sampling was introduced to the field \citep[see][for early examples]{Christensen:1998gf, Christensen:2001cr, Veitch:2009hd}, the \emph{Fisher approach} \citep{Finn:1992wt, Cutler:1994ys} was used to evaluate the sensitivity of detectors to different sources and to predict the precision with which parameters could be measured.
At its heart, this approach computes the \ac{FIM}, the noise-averaged negative Hessian of the log-likelihood (see \cref{sec:fisher}), and assumes Gaussian noise and a large-\ac{SNR} limit in which the waveform can be linearised in the parameters.
Inverting the \ac{FIM} gives an estimate of the covariance of the parameters, which \citet{Vallisneri:2007ev} argued can be interpreted either in a frequentist sense (as a lower bound on the covariance of unbiased estimators, or as the covariance of the maximum-likelihood estimator) or in a Bayesian sense (as the covariance of a multivariate normal posterior distribution under a uniform prior).
The \ac{FIM} has since largely been viewed through the Bayesian lens and remains a widely used tool for predicting the performance of future detectors \citep{Iacovelli:2022mbg,Dupletsa:2022scg}.
However, for the analysis of observational data from current-generation detectors, the method is widely considered inadequate \citep{Rodriguez:2013mla}: typical \acp{SNR} are not large, posteriors are often non-Gaussian and multimodal, and priors are not uniform.

In this work, we revisit this discussion in light of the increasing \ac{SNR} of observed signals (e.g., GW250114, which had a network \ac{SNR} of ${\sim}80$ \citep{LIGOScientific:2025rid}), investigate the utility of the Fisher approximation for inference, and propose a hybrid approach to accelerate stochastic sampling.
Specifically, we use global and multi-start optimisation to identify the modes of the posterior distribution, and construct a mixture of multivariate Gaussian distributions using the \ac{FIM} to approximate the curvature at each mode.
We show how the \ac{FIM} arises naturally as part of the Laplace approximation \citep{MacKay:2003aa}, grounding the method in a Bayesian framework.
We then view inference as a two-step process: we first rapidly construct a Laplace approximation to the posterior (a set of modes and covariances), and then refine it using stochastic sampling.
We use the resulting mixture, which we call the \emph{seeding distribution}, in four ways: directly, as an approximation to the posterior; as a proposal distribution for rejection sampling; as the initial distribution for \ac{SMC}; and as an effective prior for nested sampling via posterior repartitioning \citep{Chen:2018rep}.
This approach may be useful for current-generation detectors, but should be particularly useful for future detectors such as the Einstein Telescope \citep{Punturo:2010zz} and Cosmic Explorer \citep{Reitze:2019iox}, which will observe signals at much higher \acp{SNR} where the Fisher approximation is expected to be more accurate.

\emph{Comparison to the state of the art}~--- the use of the Fisher approximation within a hybrid inference method is not new: it is used in the \BayesStar rapid sky-localisation algorithm \citep{Singer:2015ema} and to improve the efficiency of \ac{MCMC} proposals \citep{Cornish:2005qw, Littenberg:2009bm}.
Moreover, \citet{Baka:2025yqx} recently investigated extending the Fisher approach using optimisation and prior information.
The use of posterior repartitioning for gravitational-wave nested sampling was first introduced by \citet{Prathaban:2024rmu}, while \citet{Williams:2025aar} developed \acs{ASPIRE} (\acl{ASPIRE})\acused{ASPIRE}, which initialises \ac{SMC} from a normalising flow trained on existing posterior samples, accelerating analyses that revisit an event with a new model.
The closest to our approach is the method of \citet{Prathaban:2026kft}, who combined posterior repartitioning with an initial posterior estimate from \simplepe\ \citep{Fairhurst:2023idl}; they found that \dynesty\ could fail to recover all modes and, in some cases, was slower than sampling from the prior alone.
Our method and \simplepe share a common structure: both locate the peak of the likelihood by optimisation and approximate it locally by a quadratic form (the parameter-space metric used by \simplepe is closely related to the \ac{FIM}) before using the result to inform subsequent sampling.
They differ in how this approximation is constructed: \simplepe applies the metric only to the mass and spin parameters, and obtains the distance, orientation, and sky location from physically motivated arguments based on the \acp{SNR} in each polarisation, higher multipole, and precession harmonic, whereas we apply the Laplace approximation to the full parameter space of the waveform model.
Our approach differs from the state of the art in that we use a Laplace mixture as a seeding distribution for a range of stochastic samplers, develop multi-start mode searches to handle the multimodality typical of \ac{CBC} posteriors, and implement the method within the \bilby framework \citep{Ashton:2018jfp, Romero-Shaw:2020owr}.
The implementation is provided as the \bilby plugin \texttt{bilby\_laplace} \citep{Ashton:bilbylaplace}, available at \href{https://github.com/GregoryAshton/bilby-laplace}{\texttt{github.com/GregoryAshton/bilby-laplace}}.

The remainder of this paper is structured as follows.
In \cref{sec:method}, we describe the methodology.
In \cref{sec:configuration}, we test the \ac{MAP} search, the secondary-mode search and the sampler settings in isolation.
In \cref{sec:validation}, we validate the method against state-of-the-art nested sampling and find speed-ups of up to a factor of ${\sim}4$ on gravitational-wave examples, with a trade-off between speed and accuracy.
Finally, in \cref{sec:discussion} we discuss the implications of this work and future directions.
Implementations of all the methods described here are publicly available (see the Data Availability statement at the end of this article), and these provide the complete reference for the algorithms and their settings.

\section{Methodology}
\label{sec:method}

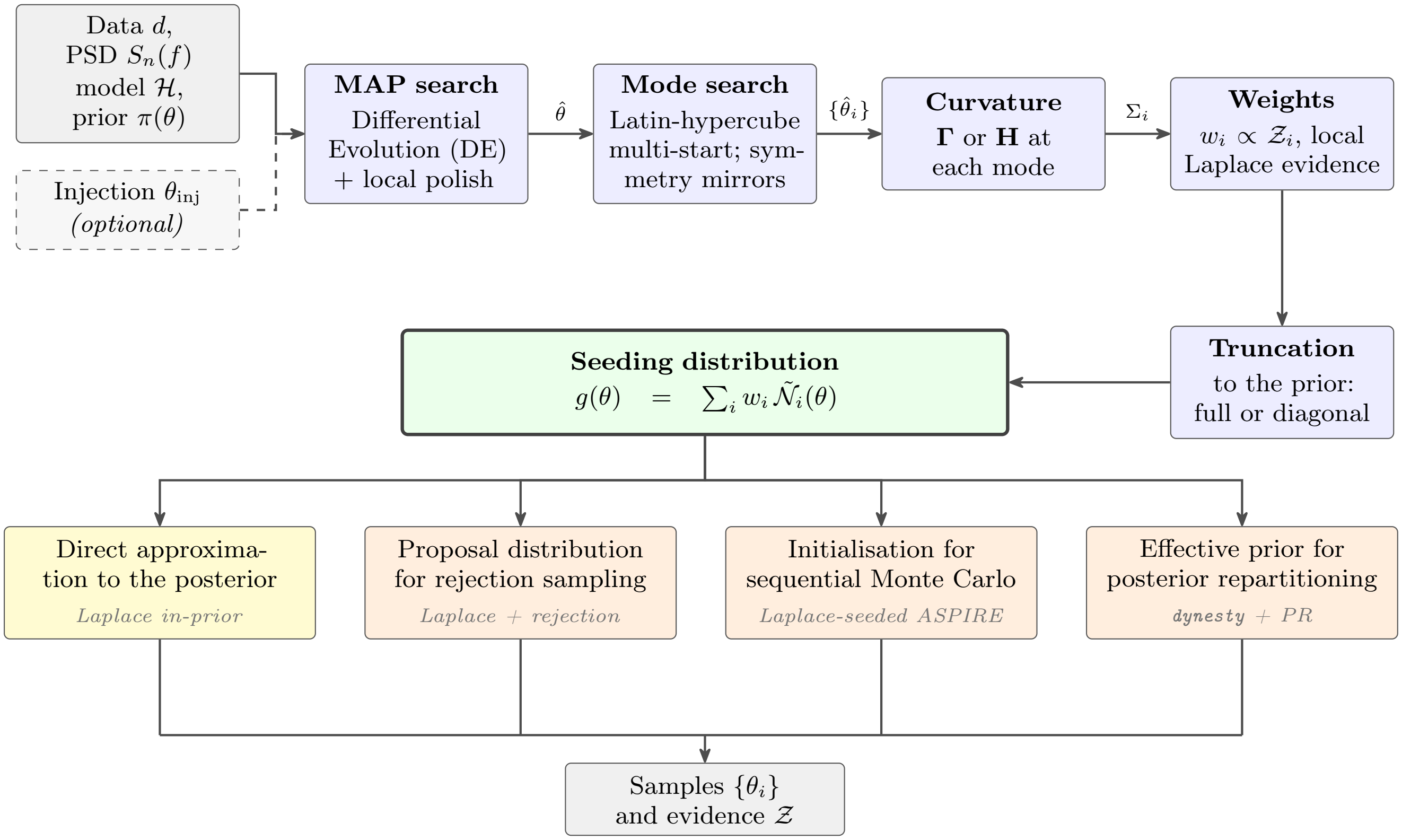
\begin{figure*}
\centering
\begin{tikzpicture}[
    font=\small,
    >={Stealth[round]},
    stage/.style={rounded corners=2pt, draw=black!65, fill=blue!7,
                  text width=25mm, align=center, inner sep=4pt, minimum height=14mm},
    io/.style={rounded corners=2pt, draw=black!65, fill=black!6,
               text width=25mm, align=center, inner sep=4pt, minimum height=9mm},
    opt/.style={io, dashed, fill=black!3},
    seed/.style={rounded corners=2pt, draw=black!75, very thick, fill=green!8,
                 text width=72mm, align=center, inner sep=5pt, minimum height=13mm},
    use/.style={rounded corners=2pt, draw=black!65, fill=orange!13,
                text width=36mm, align=center, inner sep=4pt, minimum height=14mm},
    approxuse/.style={use, fill=yellow!22},
    flow/.style={->, thick, draw=black!70},
    lab/.style={font=\scriptsize, inner sep=1.5pt, yshift=1mm},
]

\node[io]    (data)  at (1.7, 0.75) {Data \data, \acs{PSD} $S_{n}(f)$\\[1pt] model \model, prior $\prior(\parameters)$};
\node[opt]   (inj)   at (1.7,-0.95) {Injection \parametersinj\\[1pt] \emph{(optional)}};
\node[stage] (map)   at (5.3, 0.0)  {\textbf{MAP search}\\[2pt] \acf{DE} $+$ local polish};
\node[stage] (modes) at (8.9, 0.0)  {\textbf{Mode search}\\[2pt] Latin-hypercube multi-start; symmetry mirrors};
\node[stage] (cov)   at (12.5,0.0)  {\textbf{Curvature}\\[2pt] \fisher\ or \hessian\ at each mode};
\node[stage] (wts)   at (16.1,0.0)  {\textbf{Weights}\\[2pt] $w_{i} \propto \evidence_{i}$, local Laplace evidence};

\draw[flow] (data.east) -- ++(0.45,0) |- (map.west);
\draw[flow, dashed] (inj.east) -- ++(0.45,0) |- (map.west);
\draw[flow] (map)   -- node[lab, above]{\parametersstar} (modes);
\draw[flow] (modes) -- node[lab, above]{$\{\parametersstar_{i}\}$} (cov);
\draw[flow] (cov)   -- node[lab, above]{$\Sigma_{i}$} (wts);

\node[seed] (g) at (8.9,-3.1)
  {\textbf{Seeding distribution}\\[2pt]
   $g(\parameters)=\sum_{i} w_{i}\,\tilde{\N}_{i}(\parameters)$};

\node[stage] (trunc) at (16.1,-3.1) {\textbf{Truncation}\\[2pt] to the prior: full or diagonal};

\draw[flow] (wts) -- (trunc);
\draw[flow] (trunc) -- (g);

\node[approxuse] (dir)  at (2.1,-5.6)  {Direct approximation to the posterior\\[2pt] \textcolor{black!55}{\emph{\scriptsize Laplace in-prior}}};
\node[use]       (rej)  at (6.6,-5.6)  {Proposal distribution for rejection sampling\\[2pt] \textcolor{black!55}{\emph{\scriptsize Laplace + rejection}}};
\node[use]       (init) at (11.1,-5.6) {Initialisation for sequential Monte Carlo\\[2pt] \textcolor{black!55}{\emph{\scriptsize Laplace-seeded \aspire}}};
\node[use]       (rep)  at (15.6,-5.6) {Effective prior for posterior repartitioning\\[2pt] \textcolor{black!55}{\emph{\scriptsize \dynesty + PR}}};

\draw[flow] (g.south) -- ++(0,-0.55) -| (dir.north);
\draw[flow] (g.south) -- ++(0,-0.55) -| (rej.north);
\draw[flow] (g.south) -- ++(0,-0.55) -| (init.north);
\draw[flow] (g.south) -- ++(0,-0.55) -| (rep.north);

\node[io, text width=32mm] (post) at (8.9,-8.3) {Samples $\{\parameters_{i}\}$ and evidence \evidence};
\coordinate (bus) at (8.9,-7.5);
\draw[thick, draw=black!70] (dir.south)  -- (dir.south  |- bus);
\draw[thick, draw=black!70] (rej.south)  -- (rej.south  |- bus);
\draw[thick, draw=black!70] (init.south) -- (init.south |- bus);
\draw[thick, draw=black!70] (rep.south)  -- (rep.south  |- bus);
\draw[thick, draw=black!70] (dir.south |- bus) -- (rep.south |- bus);
\draw[flow] (bus) -- (post.north);

\end{tikzpicture}
\caption{
Overview of model-informed seeding; each step is described in \cref{sec:method}.
The optional injection point (dashed) is used only as a starting point for the search.
The bottom row shows the four ways in which the seeding distribution $g(\parameters)$ is used (\cref{sec:sampling}), labelled in italics with the names used in the results.
}
\label{fig:overview}
\end{figure*}

Stochastic sampling methods for approximating Bayesian posterior distributions can be thought of as a process which takes as input a distribution $g(\parameters)$ and produces as output a set of samples $\{\parameters_i\}$ which are distributed (at least asymptotically) according to the posterior distribution, \cref{eq:posterior}.
For example, a simple form of stochastic sampling is rejection sampling, in which $g(\parameters)$ is known as the proposal distribution and is used to generate candidate samples which are then accepted or rejected according to a criterion based on the posterior distribution.
The key contribution of this work is to accelerate a broad class of stochastic sampling methods by constructing an optimised $g(\parameters)$; we will refer to it as the \emph{seeding} distribution, a method-agnostic term.
\Cref{fig:overview} sets out how the seeding distribution is built and used, and the remainder of this section follows it in order.
The inputs are the standard ingredients of gravitational-wave parameter estimation \citep{LIGOScientific:2019hgc, LIGOScientific:2026ifv}: the per-detector interferometric strain data and estimated noise \ac{PSD}, a signal model \model, and the prior.
Optionally, our proposed algorithm also takes as input \parametersinj, the parameters used to simulate the data, usually referred to as the \emph{injection} parameters.
This enters only as an initialisation point for the search, and we envision it being used in simulation studies where the cost of a blind search is not itself the object of study.
A local search started from \parametersinj\ is roughly a hundred times cheaper than a blind search, but it does not establish that the point found is the global maximum.
We therefore do not use it in the coverage tests of \cref{sec:pp}, where it would leave only the sampling stage under test.
In observational analyses the injection is unknown; the parameters reported by the search that identified the signal could play the same role, but we do not consider this here.

\subsection{Locating the maximum \emph{a posteriori} point}
\label{sec:map}
The first stage of the construction is a global search for the \ac{MAP} point \parametersstar.
For convenience, we work with the unnormalised posterior distribution and its negative logarithm, referred to as the potential:
\begin{align}
    \unnorm(\parameters|\data, \model) &:= \likelihood(\data|\parameters, \model) \prior(\parameters|\model)\,,
    \label{eq:unnorm} \\
    \potential(\parameters) &:= -\ln \unnorm(\parameters|\data, \model)\,,
    \label{eq:potential}
\end{align}
so that \parametersstar\ is the minimum of \potential\ over the support of the prior.
Within \bilby, some parameters can be declared as periodic \citep{Romero-Shaw:2020owr}, e.g. phase parameters which are periodic in $2\pi$.
For these parameters, we also wrap them onto their period before the potential is evaluated, so that the optimiser does not see an artificial boundary in the middle of a mode.

This stage is critical: failure to identify the \ac{MAP} most likely leads to catastrophic failure of the whole method.
In initial testing, we found that local gradient-based optimisation started from prior draws is not adequate because the potential is strongly multimodal.
This difficulty has previously motivated the use of population-based optimisers in gravitational-wave data analysis, such as genetic algorithms \citep{Crowder:2006wh} and particle swarm optimisation \citep{Wang:2010jma}, the latter recently applied to searches for stellar-origin binaries in data from the \ac{LISA} \citep{Bandopadhyay:2023gkb, Fu:2024cpu}.
Therefore, we use \acf{DE} \citep{Storn:1997uea}, a population-based stochastic optimiser which requires no derivatives of the potential and searches the whole prior volume.
\ac{DE} is already used in gravitational-wave inference: it refines template parameters in the PyCBC Live low-latency search \citep{Trevor:2026dgs}, and it finds the maximum-likelihood fiducial parameters for the relative-binning likelihood in \bilby \citep{Krishna:2023bug}.
We base our algorithm around the \scipy\ \citep{Virtanen:2019joe} implementation and inherit the default settings (version 1.18.0 for the results presented here) except in three respects.
First, we declare convergence once the standard deviation of the potential across the population falls below an absolute threshold of 1 nat; the default criterion is relative to the mean of the potential, which is dominated by an additive constant unrelated to the structure of the posterior.
Second, we evaluate the whole population in a single batch at each generation, so that its likelihood evaluations can be parallelised.
Third, \scipy's own final refinement relies on finite-difference gradients, which are unreliable for parameters whose scales span several orders of magnitude, so we replace it with a \emph{local polish}: a local optimisation by the derivative-free Nelder--Mead simplex method \citep{Nelder:1965}, using \scipy's implementation with its default settings, started from the best member of the final population and constrained to the prior bounds.
\ac{DE} locates the basin containing the minimum, and the polish descends to the bottom of that basin.
We then repeat the whole search from \nrestarts independent populations and retain the best result.

If \parametersinj is provided, the \ac{DE} step and its restarts are skipped, and \parametersstar\ is found by a single Nelder--Mead optimisation started from \parametersinj, with the same wrapping of periodic parameters and the same prior bounds as the local polish.

When applying the \ac{MAP} search in practice, with \texttt{IMRPhenomXPHM\_SpinTaylor} \citep{Pratten:2020ceb,Colleoni:2024knd}, the search occasionally ends on a narrow spike in the log-posterior, a numerical artefact of the precession prescription, whose curvature then describes the spike rather than the posterior.
We therefore apply a guard to every \ac{MAP} search which discards such candidates; the identification of the spikes and the guard are described in \cref{sec:spikes}.

\subsection{Searching for secondary modes}
\label{sec:modes}
Gravitational-wave posterior distributions are often multimodal, especially in the extrinsic parameters describing the source's sky location and orientation.
The \ac{MAP} search can only identify \parametersstar of a single mode and so an approximation built on it alone is inadequate.
We therefore search for further modes $\{\parametersstar_i\}$ with the aim of building the seeding distribution $g(\parameters)$ as a mixture of multivariate Gaussian distributions, one for each mode.

We use two complementary searches to propose candidates.
The first is stochastic: we draw a Latin hypercube over the prior (optionally restricted to a subspace of the parameters with the remainder held at \parametersstar), evaluate the potential at each point, and polish those with the lowest potential by local optimisation in the full parameter space.
Latin hypercube sampling is used in preference to independent prior draws because it guarantees coverage of each one-dimensional marginal, which for a fixed and modest number of evaluations gives a more uniform exploration of the prior volume.
The second is deterministic: many of the modes of a gravitational-wave posterior are the images of the dominant mode under known symmetries of the signal model \citep[see, e.g.][]{Roulet:2022kot}.
We therefore construct the corresponding mirror points directly from \parametersstar\ and the other candidates, and retain those whose potential lies close to that of their source.
Because a mirror is a translation of its source, it inherits the source's covariance, so each costs a single likelihood evaluation; these candidates locate modes which are narrow in the full parameter space and which the stochastic search finds only unreliably.
Finally, we also implement two optional variants: a \emph{full-space multi-start} search, which polishes every point of a Latin hypercube before ranking them, and \emph{symmetric polishing}, which starts a local optimisation from each mirror point that lies too far below its source.

Candidates from both searches are then ranked by their log-posterior, highest first.
We accept them in order, discarding any candidate which lies within a fixed number of standard deviations of an already-accepted mode, so that the same basin is not represented twice; the scale for this comparison is taken from the per-parameter standard deviations of the dominant mode.
The search terminates once a specified number of modes has been accepted or the pool is exhausted.
We stress that this procedure offers no guarantee of having found every mode, and that a mode which is missed is simply absent from the seeding distribution; this can potentially be saved by some of the sampling methods (see \cref{sec:sampling}) but typically at a cost of reduced efficiency.

\subsection{Curvature from Laplace's approximation}
\label{sec:laplace}
With the set of local \ac{MAP} points $\{\parametersstar_i\}$, we now need to evaluate the curvature of the log-posterior at each mode to construct the mixture model.
To do this, we utilise Laplace's approximation \citep{MacKay:2003aa}, which approximates the posterior distribution as a multivariate Gaussian distribution with mean \parametersstar and covariance given by the inverse of the Hessian of the potential at that point.

To derive the Laplace approximation, one can start from the unnormalised posterior distribution of \cref{eq:unnorm}; by Bayes' theorem, the joint distribution of the parameters and data is equal to the unnormalised posterior distribution and the product of the evidence and the posterior distribution:
\begin{align}
    \posterior(\parameters, \data| \model) = \unnorm(\parameters | \data, \model) = \posterior(\parameters|\data, \model) \evidence(\data|\model)\,.
    \label{eq:joint}
\end{align}

Laplace's approximation then proceeds by approximating the joint distribution as an unnormalised multivariate Gaussian distribution, which is most easily seen through the potential of \cref{eq:potential}.
Given a local minimum \parametersstar of the potential, equivalently a \ac{MAP} point, the gradient vanishes and a Taylor expansion to second order gives 
\begin{align}
    \potential(\parameters) \approx \potential(\parametersstar) + \frac{1}{2} (\parameters - \parametersstar)^{\transpose} \hessian (\parameters - \parametersstar)\,,
\end{align}
where $\hessian$ is the Hessian matrix of second derivatives of the potential evaluated at \parametersstar:
\begin{align}
    \hessian_{ab} = \left. \frac{\partial^2 \potential}{\partial \theta^{a} \partial \theta^{b}} \right|_{\parameters = \parametersstar}\,.
    \label{eq:hessian}
\end{align}
Then exponentiating the potential and factoring into a constant term and \parameters-dependent term, we obtain the Laplace approximation to the unnormalised posterior distribution:
\begin{align}
    \unnorm(\parameters|\data,\model) &\simeq
    e^{-\potential(\parametersstar)}
    \exp\left[-\tfrac{1}{2}(\parameters-\parametersstar)^{\transpose}
    \hessian(\parameters-\parametersstar)\right] \\
    &= \underbrace{e^{-\potential(\parametersstar)}
      \sqrt{(2\pi)^{\nparameters}\left|\hessian^{-1}\right|}}_{\text{independent of }\parameters}
    \;\times\;
    \underbrace{\N\!\left(\parameters;\parametersstar,\hessian^{-1}\right)}_{
      \text{normalised over }\parameters}\,.
    \label{eq:factorised}
\end{align}
Comparison with \cref{eq:joint} reveals that the posterior distribution is a normalised multivariate Gaussian distribution with mean \parametersstar and covariance matrix $\hessian^{-1}$:
\begin{align}
    \posterior(\parameters|\data, \model) \approx \N\!\left(\parameters;\parametersstar,\hessian^{-1}\right)\,,
    \label{eq:laplace}
\end{align}
while the evidence is given by
\begin{align}
    \evidence(\data|\model) \approx \unnorm(\parametersstar|\data, \model) \sqrt{(2\pi)^{\nparameters} \left|\hessian^{-1}\right|}\,,
\end{align}
where \nparameters is the number of parameters.

The Laplace approximation is built entirely around the local maximum \parametersstar; it is therefore symmetric and unimodal by construction.
Provided the Hessian can be evaluated, the Laplace approximation is a fast and efficient way to approximate the posterior distribution and evidence.
For a general likelihood and prior, the Hessian can be evaluated numerically using finite difference or automatic differentiation methods.
However, for the specific case of a Gaussian likelihood, the Hessian can be further simplified.

\subsection{Curvature for stationary Gaussian noise}
\label{sec:fisher}
We consider data \data\ consisting of a signal $\signal(\parameters)$ and additive noise \noise, the latter a zero-mean Gaussian process which we take to be stationary over the analysis segment with known one-sided \ac{PSD} $S_{n}(f)$ (which in practice must be estimated from the data).
The noise covariance then induces the noise-weighted inner product
\begin{align}
    \inner{x}{y} := 4\,\mathrm{Re} \int_{0}^{\infty} \frac{\tilde{x}^{*}(f)\,\tilde{y}(f)}{S_{n}(f)}\,\mathrm{d}f\,,
    \label{eq:inner-product}
\end{align}
where a tilde denotes the Fourier transform, so that, up to an additive constant,
\begin{align}
    \ln\likelihood
    &= -\tfrac{1}{2}\inner{\data - \signal(\parameters)}{\data - \signal(\parameters)}\,.
\end{align}
What follows uses only the bilinearity of \cref{eq:inner-product} and the vanishing first moment of the noise, so it applies unchanged to any zero-mean Gaussian noise covariance, \cref{eq:inner-product} being the form relevant for gravitational-wave data.

Writing $\partial_a \equiv \partial/\partial\theta^{a}$, we then have
\begin{align}
    \partial_a\partial_b \ln\likelihood
    &= \inner{\data - \signal(\parameters)}{\partial_a\partial_b \signal(\parameters)}
    - \inner{\partial_a \signal(\parameters)}{\partial_b \signal(\parameters)}\,,
\end{align}
such that the Hessian, \cref{eq:hessian}, can be evaluated from
\begin{equation}
\begin{split}
    \hessian_{ab} = \bigl[
    &\inner{\partial_a \signal(\parameters)}{\partial_b \signal(\parameters)} \\
    & - \inner{\data - \signal(\parameters)}{\partial_a\partial_b \signal(\parameters)} \\
    & - \partial_a\partial_b \ln\prior
    \bigr]_{\parameters = \parametersstar}\,.
\end{split}
    \label{eq:exact-hessian}
\end{equation}

So far this is an exact expression for the Laplace approximation Hessian, but it may be simplified by noting that, in the large \ac{SNR} limit, the second term becomes negligible relative to the first.
There are two distinct routes to this conclusion.

The first is to average over noise realisations of data generated at parameters $\parameters$, for which $\data - \signal(\parameters) = \noise$ and $\noiseavg[\parameters]{\noise} = 0$, so that the term vanishes in expectation and
\begin{align}
    \fisher_{ab}(\parameters) :=
    \noiseavg[\parameters]{-\partial_a\partial_b \ln\likelihood}
    = \inner{\partial_a \signal}{\partial_b \signal}\,,
    \label{eq:fisher}
\end{align}
which is the \ac{FIM}.
The second is to linearise the signal about the true parameters $\parameterstrue$, which sets $\partial_a\partial_b \signal = 0$ identically and so removes the term for every realisation rather than merely on average, yielding, again, the \ac{FIM}, \cref{eq:fisher}.
Evaluated at \parameterstrue, the \ac{FIM} depends only on the signal model and the noise \ac{PSD}, which is the sense in which it is used for forecasting; since \parameterstrue\ is unknown when analysing observational data, we instead evaluate it at the mode, so that it retains a dependence on the noise realisation through \parametersstar.
\Citet{Vallisneri:2007ev} showed that no single \ac{SNR} threshold establishes the validity of \cref{eq:fisher}.

Starting from the Laplace approximation of \cref{eq:laplace}, rather than assuming a constant prior, the prior-informed Hessian at mode $i$ is then approximated by
\begin{align}
    \hessian_{i,ab} \approx \left[ \fisher_{ab}(\parameters) - \partial_a\partial_b\ln\prior\right]_{\parameters=\parametersstar_i}\,.
    \label{eq:prior-informed-hessian}
\end{align}
For a Gaussian prior this reduces to the additive prior-precision term already given by \citet{Vallisneri:2007ev}; the form above holds for an arbitrary prior.

We evaluate the curvature in one of two ways (\cref{tab:settings}).
For the gravitational-wave examples, we use \cref{eq:prior-informed-hessian}, with \fisher computed from finite-difference derivatives of the waveform projected onto each detector.
For the low-dimensional examples, we instead evaluate \cref{eq:hessian} directly, by finite differences of the log-posterior in the unit hypercube on which the prior is uniform.
Where the likelihood is analytically marginalised over some parameters (e.g., the phase, time, or distance), we reinstate them in \fisher and then marginalise them out of the resulting covariance, which accounts for their correlations with the sampled parameters rather than fixing them at the mode.
In both cases, we bound the covariance so that no parameter is wider than its prior: a direction which the data leave unconstrained, or in which the finite-difference estimate is noisy, otherwise inverts to an arbitrarily large variance.
For the waveform \ac{FIM}, we cap the marginal variance of each parameter at that of its prior, leaving the correlations unchanged; for the unit-hypercube Hessian, we floor its eigenvalues at the precision of a uniform distribution on the unit interval.
We write $\Sigma_i$ for the resulting covariance of mode $i$, which equals $\hessian_i^{-1}$ wherever the bound is inactive.

\subsection{Mode weights}
\label{sec:weights}
Each mode consists of a location $\parametersstar_i$ and covariance $\Sigma_i$ (\cref{sec:fisher}); we build the seeding distribution as a mixture of these modes, but we must first assign them weights $w_i$.
We implement two options: equal weighting, which gives every accepted mode the same weight, and weighting by the local Laplace evidence, which accounts for the relative volume of each mode.
While the former is simple and robust, it gives a mode holding little posterior mass as many draws as the dominant one; we therefore prefer the latter.
Namely, we apply \cref{eq:factorised} locally to each mode giving its Laplace evidence:
\begin{align}
    \ln\evidence_i = \ln\unnorm(\parametersstar_i|\data,\model)
    + \frac{\nparameters}{2}\ln(2\pi) + \frac{1}{2}\ln\left|\Sigma_i\right|\,,
    \label{eq:mode-evidence}
\end{align}
in which the second term is common to every mode and cancels on normalisation, while the third supplies the volume factor required to prevent a narrow tall mode from dominating a broad shallow one.
We note however that the Laplace evidence is approximate and non-Gaussian modes or misestimation of the covariance can lead to mis-weighting and inefficient sampling.

Normalising over the modes, we set weights $w_i \propto \evidence_i$ such that the seeding distribution is the mixture
\begin{align}
    g(\parameters) = \sum_i w_i \,\N\!\left(\parameters;\parametersstar_i, \Sigma_i\right)\,,
    \label{eq:seeding}
\end{align}
which is \cref{eq:laplace} generalised from one mode to many and reduces to it when a single mode is retained.
Modes with a weight below $10^{-3}$ are dropped, although the mode of largest weight is always kept: such a mode would receive only a handful of draws, too few to sample it, and would contribute only stray points.

\subsection{Prior truncation}
\label{sec:truncation}
\begin{figure}
    \centering
    \includegraphics[width=\columnwidth]{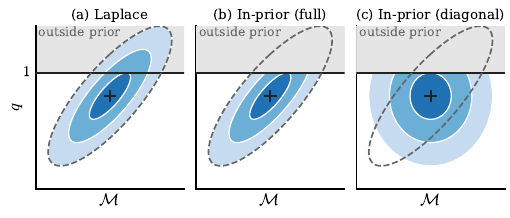}
    \caption{
    Illustration of the in-prior truncation of a component whose mode (cross) lies near the bound $q \le 1$.
    (a) The untruncated Gaussian places a fraction of its draws outside the prior.
    (b) The full truncation retains the correlation.
    (c) The diagonal truncation discards it.
    In each panel, the dashed curve is the outer contour of the untruncated Gaussian.
    }
    \label{fig:truncation}
\end{figure}

\Cref{eq:seeding} can be used as a seeding distribution, but retaining the prior curvature in \cref{eq:prior-informed-hessian} does not capture every effect of the prior.
A prior which is uniform over a bounded region contributes no curvature in the interior but truncates the support, and the covariance $\Sigma_i$ carries no knowledge of where the support ends.
Therefore, for a parameter whose prior is narrow compared with the curvature scale, or whose mode lies close to a bound, a Gaussian component places an appreciable fraction of its draws outside the prior, where they will be immediately rejected, and may describe the posterior poorly even at high \ac{SNR}.
We therefore truncate each component to the prior bounds, replacing it by a component $\tilde{\N}_i(\parameters)$ which is normalised over the support of the prior, so that the seeding distribution becomes
\begin{align}
    g(\parameters) = \sum_i w_i \,\tilde{\N}_i(\parameters)\,;
    \label{eq:seeding-truncated}
\end{align}
we refer to this as \emph{Laplace in-prior} seeding.
We implement two ways to truncate the components to the prior, which trade fidelity to $\Sigma_i$ against cost and simplicity; \cref{fig:truncation} illustrates both.

First, the most faithful is to take $\tilde{\N}_i$ to be $\N(\parametersstar_i, \Sigma_i)$ truncated, as a multivariate Gaussian, to the prior bounds; we refer to this as \emph{Laplace in-prior (full)}.
Drawing from it is straightforward, by discarding draws which fall outside the bounds, but evaluating its density, as some downstream sampling methods require, needs its normalisation: the mass of $\N(\parametersstar_i, \Sigma_i)$ within the bounds.
Beyond a few dimensions this has no closed form and must be estimated numerically, and any error in the estimate enters the weights and the evidence directly.

Second, the simpler approach is to take $\tilde{\N}_i$ to be a product of one-dimensional Gaussians with widths $\sigma_{i,a} = (\Sigma_i)_{aa}^{1/2}$, each truncated to its own prior bounds, for which drawing and the density are both exact and cheap. We refer to this as \emph{Laplace in-prior (diagonal)}.
The cost is that correlations between parameters are discarded: the product spans the same range as the posterior in every parameter, but does not follow it along correlated directions.
This matters less than it sounds for seeding a sampler, since a covariance that is too narrow is more harmful than one that is uncorrelated.
Later, in the results sections, we find that the diagonal truncation is the more efficient seed when $\Sigma_i$ underestimates the width of the posterior.
The full covariance $\Sigma_i$ is nonetheless used in \cref{eq:mode-evidence} to weight the modes.

In this work we predominantly use diagonal truncation for seeding, but the full method is available in \texttt{bilby\_laplace} and, because it retains the correlations in $\Sigma_i$, provides an improved estimate if the Laplace approximation is to be used directly as a posterior approximation.

The prior may also bound quantities derived from several sampled parameters, such as the component masses when sampling in chirp mass and mass ratio; we refer to these as \emph{constraint priors}.
They cannot be imposed one parameter at a time, so draws which violate them are instead discarded by each of the sampling methods below.

\subsection{Posterior sampling}
\label{sec:sampling}
Having obtained our seeding distribution, \cref{eq:seeding-truncated}, we now describe the four ways in which we use it to generate posterior samples.
They differ in what they do about the discrepancy between $g(\parameters)$ and $\posterior(\parameters|\data,\model)$: the first ignores it, while the remaining three each correct for it asymptotically, so that the seeding distribution affects only the cost of the analysis and not its result.
This is the sense in which model-informed seeding is safer to use than the Laplace approximation alone: an inaccurate $g(\parameters)$ makes the last three methods slower, but the resulting bias is corrected.
This guarantee holds only asymptotically; in a run of finite length, a poor seeding distribution can still bias the result, for example by under-sampling a mode which it misses.

\subsubsection{In-prior sampling}
\label{sec:inprior}
The simplest use of the seeding distribution is to draw from it directly, discard any draw which violates a constraint prior, and evaluate the likelihood at those that remain.
The resulting samples are distributed according to $g(\parameters)$ restricted to the prior, and are therefore only as accurate as the Laplace approximation itself; no correction is applied.
Its cost is one likelihood evaluation per retained sample, which is the smallest cost at which a set of samples can be produced at all.
We include it for two reasons.
First, it is a diagnostic tool, which isolates the quality of the seeding distribution: any discrepancy between an in-prior result and a reference posterior is attributable to the approximation of \cref{eq:seeding-truncated} alone, with no sampler behaviour mixed in.
Second, it is the natural extension of \ac{FIM} forecasting, which predicts a covariance but no samples: drawing in this way produces a sample set which respects the bounds of the prior and the multimodality of the posterior, both of which a single Fisher ellipse omits, at a cost far below that of a full analysis.
\Citet{Baka:2025yqx}, following \citet{Dupletsa:2022scg}, similarly drew samples from a Fisher approximation truncated to the prior bounds, though about a single mode.
We implement and test in-prior sampling with both the full and diagonal truncations of \cref{sec:truncation}.

\subsubsection{Rejection sampling}
\label{sec:rejection}
Rejection sampling converts draws from $g(\parameters)$ into exact posterior samples.
A draw $\parameters \sim g$ is accepted with probability
\begin{align}
    \alpha(\parameters) = \frac{\unnorm(\parameters|\data,\model)}{M g(\parameters)}\,,
    \label{eq:acceptance}
\end{align}
where $M \geq \sup_{\parameters} \unnorm(\parameters|\data,\model)/g(\parameters)$.
We establish $M$ from a pre-scan batch of draws taken before any sample is accepted, rather than updating it as the run proceeds, since a bound raised mid-run would invalidate the samples already accepted under the smaller value.
The accepted samples are independent and identically distributed, are equally weighted, and the mean acceptance probability provides an estimate of the evidence together with a direct measure of how well $g(\parameters)$ matches the posterior.

The same draws could instead be retained with importance weights $\unnorm(\parameters)/g(\parameters)$, which wastes no likelihood evaluations.
We do not do this, because the output is then a weighted sample set whose effective size is smaller than its nominal size and whose members are not independent.
Independent, equally weighted samples are what downstream analyses consume, and rejection sampling produces them directly.

\subsubsection{Sequential Monte Carlo}
\label{sec:smc}
Rejection sampling is efficient only where $g(\parameters)$ is close to the posterior everywhere, since a single poorly-matched region inflates $M$ and depresses the acceptance rate throughout.
\ac{SMC} relaxes this by bridging from the seeding distribution to the posterior through a sequence of intermediate distributions, moving a population of particles between them by \ac{MCMC} and resampling to control the spread of their weights.
We use the \aspire\ implementation \citep{Williams:2025aar}, through its \bilby interface \texttt{aspire-bilby}, using version 0.1.0 of both \citep{Williams:aspire,Williams:aspirebilby} for the results presented here.
\ac{SMC} has been validated for gravitational-wave inference by \citet{Williams:2025szm} using persistent sampling as implemented in \pocomc \citep{Karamanis:2022ksp}, with its own preconditioning and sampler configuration; we validate the \aspire\ implementation and settings used here against \dynesty\ in \cref{sec:settings,sec:validation}.
In \aspire, the tempered target at inverse temperature $\beta$ is
\begin{align}
    \ln \posterior_{\beta}(\parameters) = (1-\beta)\ln q(\parameters)
    + \beta \ln \unnorm(\parameters|\data,\model)\,,
    \label{eq:tempered}
\end{align}
so that $\beta = 0$ recovers the reference distribution $q$ and $\beta = 1$ the posterior, with the schedule of intermediate $\beta$ chosen adaptively.
At each iteration, the next value of $\beta$ is found by bisection, to a tolerance of $10^{-6}$, as the largest value for which the effective sample size of the incremental importance weights, as a fraction of the number of particles, is at least a target efficiency $e(\beta) = e_0 + (e_1 - e_0)\beta^{r}$ evaluated at the current $\beta$; no minimum or maximum step in $\beta$ is imposed.
For every analysis, we use $e_0 = 0.5$, $e_1 = 0.8$ and $r = 0.5$ (\cref{tab:settings}), so that the steps become more conservative as $\beta$ approaches 1.
After each step, the particles are resampled multinomially, irrespective of their effective sample size, and each is then moved by a fixed number of \ac{MCMC} steps using the $t$-preconditioned Crank--Nicolson proposal \citep{Grumitt:2024tpcn} as implemented in \minipcn \citep{Williams:minipcn} (version 0.2.0a3), which fits a multivariate Student-t distribution to the particle population at the start of each mutation stage and adapts the proposal's correlation parameter towards a target acceptance rate of 0.234.
The mutations are performed in the coordinates in which the prior is defined, with periodic parameters wrapped onto their period and no further transformation; proposals outside the prior support are rejected through the prior.
Here the seeding distribution enters twice: the initial particle cloud is drawn from $g(\parameters)$, and $q$ is a normalising flow trained on draws from $g(\parameters)$, so that the seeding distribution is present in the target at every temperature rather than merely at the start.
We also implemented the option of using the mixture $g(\parameters)$ itself as $q$, avoiding the flow training, but found it performed substantially worse on the BBH HLV example (\cref{sec:simulated}), because the mixture is unbounded and does not respect the periodic parameters, whereas the trained flow does via its bounded and periodic transforms in \aspire.
We also implement the option of drawing a fraction of the initial cloud from the prior instead, which provides particles in regions the mode search did not reach; at $\beta \simeq 0$ such a particle carries ordinary weight and need only survive until the schedule reaches a temperature at which its mode is resolved.
This option is available in \texttt{bilby\_laplace}, but we do not use it for the results presented here.

\subsubsection{Posterior repartitioning}
\label{sec:pr}
The final use applies the seeding distribution to nested sampling, which is a standard tool for gravitational-wave inference and which we use through the \dynesty\ implementation \citep{Speagle:2019ivv} as interfaced by \bilby\ \citep{Ashton:2018jfp}.
Nested sampling begins with live points drawn from the prior and contracts towards the bulk of the likelihood, so for a high-\ac{SNR} signal, whose posterior occupies a very small fraction of the prior volume, the great majority of the computation is spent on this contraction.
Posterior repartitioning \citep{Chen:2018rep, Chen:2023rep}, applied to gravitational-wave nested sampling by \citet{Prathaban:2024rmu}, exploits the fact that the split of the unnormalised posterior into a likelihood and a prior is not unique.
Writing
\begin{align}
    \unnorm(\parameters|\data,\model)
    = \underbrace{\frac{\likelihood(\data|\parameters,\model)\prior(\parameters|\model)}{g(\parameters)}}_{\tilde{\likelihood}(\parameters)}
    \times \underbrace{\vphantom{\frac{\likelihood}{g}} g(\parameters)}_{\tilde{\prior}(\parameters)}\,,
    \label{eq:repartition}
\end{align}
we run nested sampling with $\tilde{\prior}$ as its prior and $\tilde{\likelihood}$ as its likelihood.
The product is unchanged, so the posterior and the evidence are unchanged; what changes is that the live points start already distributed according to the seeding distribution, and the volume the algorithm must contract through is correspondingly smaller.
\citet{Prathaban:2026kft} took this approach with $g(\parameters)$ constructed from \simplepe\ estimates \citep{Fairhurst:2023idl}; here, $\tilde{\prior}$ is instead built from the Laplace seeding distribution.
We use a simple construction: since \dynesty draws from its prior by transforming each coordinate of the unit hypercube independently, we take $\tilde{\prior}$ to be a product of one-dimensional distributions, which the mixture of \cref{eq:seeding-truncated} is not.
We therefore use only the component at the \ac{MAP} point, with the diagonal truncation of \cref{sec:truncation}, and widen each of its widths by a factor of two (\cref{tab:settings}) so that $\tilde{\prior}$ is wider than the posterior in every direction.
Periodic parameters keep their original prior, since a truncated Gaussian cannot follow a posterior across the wrap; the mirror modes of \cref{sec:modes} differ from the \ac{MAP} mode only in such parameters, so they lose nothing.
The requirement is that $\tilde{\prior}$ be non-zero wherever the prior is, since $\tilde{\likelihood}$ is otherwise unbounded; each truncated Gaussian has support over the full prior range of its parameter, so this holds, and a secondary mode which $\tilde{\prior}$ does not cover in principle costs only iterations.
The full mixture could be used instead by writing it as a sequence of one-dimensional conditional distributions, but we leave this to future work.

\section{Configuration}
\label{sec:configuration}
In this section, we describe the set of studies we performed to choose the configuration of the \ac{MAP} search, the secondary-mode search, and the \ac{SMC} sampler.

\subsection{Locating the maximum \emph{a posteriori} point}
\label{sec:map-validation}
To test how closely the search of \cref{sec:map} approaches the \ac{MAP} point, we simulated 100 precessing \ac{BBH} signals observed by the \ac{LHO}, \ac{LLO} and Virgo network (drawing the properties and settings from the BBH HLV example discussed later in \cref{sec:simulated}).
The prior draws are set up such that the optimal network \ac{SNR} ranges from 12 to 40.
We scored the first 100 simulations to complete, excluding two in which the waveform model hangs at a point the search converged onto.
For each signal, we ran eight searches with $\nrestarts=4$, giving 32 independent restarts per signal.
Since the true \ac{MAP} point is unknown, we polish the result of each search and the injected parameters with repeated Nelder--Mead optimisations and take the highest point as the reference, with log-posterior $\ln\posterior_{\rm ref}$.
We then define the shortfall of a search whose result has log-posterior $\ln\posterior_{\rm MAP}$ as $\ln\posterior_{\rm ref} - \ln\posterior_{\rm MAP}$, a lower bound on the shortfall relative to the true \ac{MAP} point.
We score a search with \nrestarts restarts as the best of \nrestarts of the 32, with the distribution averaged over all such groups and restarts rejected by the spike guard counted as misses.

\begin{figure}
    \centering
    \includegraphics[width=\columnwidth]{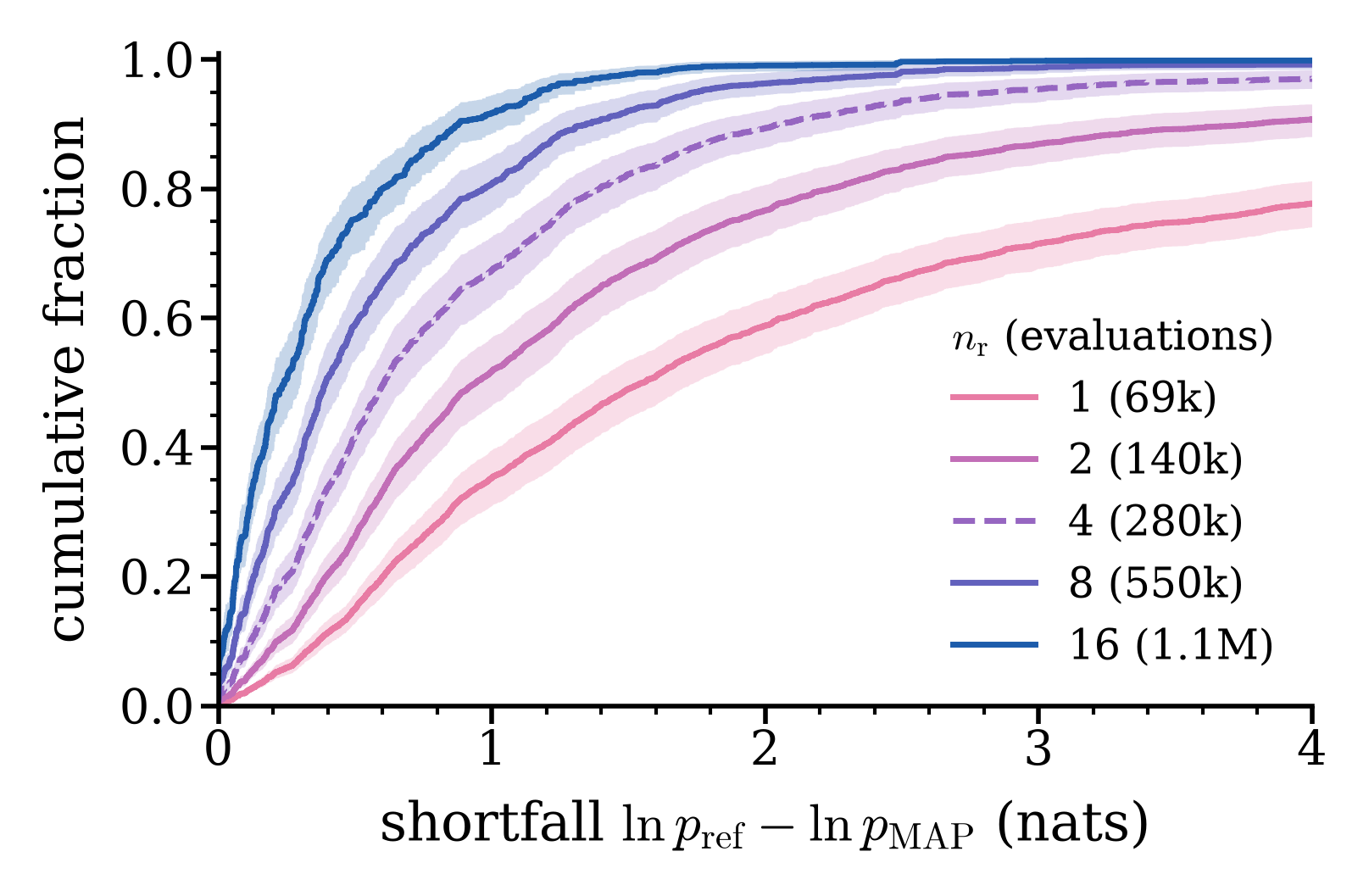}
    \caption{
    Cumulative distribution of the \ac{MAP} search's shortfall over 100 simulated precessing \ac{BBH} signals, for searches with \nrestarts restarts (the default, $\nrestarts=4$, dashed); the legend gives the mean number of likelihood evaluations.
    Bands are 5--95\% bootstrap ranges over signals.
    Shortfalls beyond 4 nats are off-axis but counted.
    }
    \label{fig:map-shortfall}
\end{figure}

\Cref{fig:map-shortfall} shows the results.
To give a sense of scale, if the posterior is approximately Gaussian in its $d$ sampled parameters, a typical posterior sample lies ${\sim}d/2$ nats below the peak.
Hence, for the 13 parameters sampled here, a typical shortfall would be ${\sim}6.5$ nats.
This supports the default, $\nrestarts=4$, for which the median shortfall is 0.6 nats and the 90th percentile is 2.0 nats, so 90\% of searches end higher than all but ${\sim}1\%$ of the posterior.
Further increasing \nrestarts\ gives diminishing returns at a linearly growing cost: doubling to $\nrestarts=8$ reduces the 90th percentile only to 1.3 nats.
Most of the remaining shortfall arises because a search ends on a neighbouring local maximum, not because it stops short of the top of the one it finds.\footnote{At $\nrestarts=4$, 21 of the 800 searches fall short by more than 4 nats, from only four signals; six are from one simulation which has a sky position just $1.6^\circ$ from the celestial pole.
Here, the posterior has a ridge in azimuth and polarisation angle narrower than $\delta$ in \cref{eq:spike-guard}.
As a result, the spike guard flags it, rejecting a genuine maximum: this suggests improvements to the guard are required in future development.}
Overall this suggests that $\nrestarts=4$ is a good compromise that generally tends to find a sufficient approximation of the \ac{MAP} point. We further validate this later in \cref{sec:pp} by performing a \ac{PP} test which builds the seeding distribution from a blind \ac{MAP} search.

\subsection{Searching for secondary modes}
\label{sec:mode-validation}
To test the secondary-mode search of \cref{sec:modes}, we simulated a further set of precessing \ac{BBH} signals from the same prior, with the luminosity distance restricted to 100--1000\,Mpc so that the median network \ac{SNR} is ${\sim}30$.
As in \cref{sec:map-validation}, we had to over-cover the simulations where the waveform model was found to hang; six signals were excluded for this reason, leaving 100.
Each signal is analysed in a paired ablation study: every arm starts from the identical \ac{MAP} point and enables a single secondary-mode search with up to five modes, while two further arms enable every search with up to five and fifteen modes, and a final arm uses the configuration of the examples of \cref{sec:validation} (hypercube and symmetry searches with up to three modes).
Components closer than three standard deviations to one already in the mixture are rejected in every arm, and the two optional variants of \cref{sec:modes} are each tested in their own arm and enabled in the combined arms.

Because each signal is drawn from the prior, its injected parameters are, marginally over the population, a draw from the posterior, and so should lie close to some component of a seeding distribution which approximates it.
We therefore measure the distance from the injected parameters to the nearest component, taking the largest over parameters of the offset in units of that component's standard deviation, and count a signal as recovered when this is below 3; for a calibrated Gaussian in 13 dimensions, 97\% of draws satisfy this.

\begin{figure}
    \centering
    \includegraphics[width=\columnwidth]{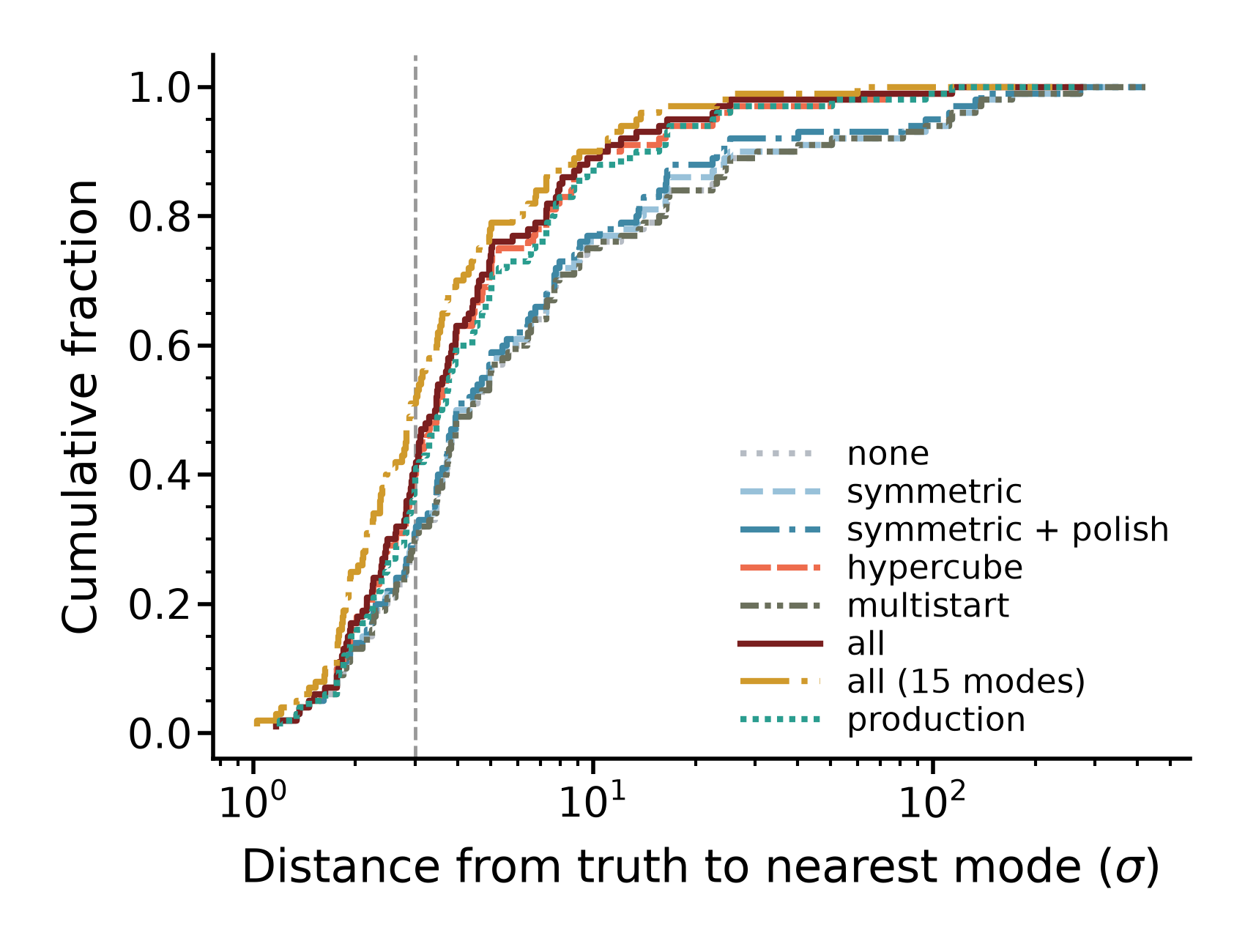}
    \caption{
    Cumulative distribution, over 100 simulated precessing \ac{BBH} signals, of the distance from the injected parameters to the nearest component of the seeding distribution, in units of that component's standard deviation.
    Each curve is one arm of the paired ablation: ``none'' retains the \ac{MAP} mode alone; ``symmetric'', ``symmetric + polish'', ``hypercube'' and ``multistart'' each enable a single search; ``all'' and ``all (15 modes)'' enable every search with up to five and fifteen modes; and ``production'' enables the hypercube and symmetry searches with up to three modes, as in the examples.
    The vertical grey line marks $3\sigma$.
    }
    \label{fig:mode-recall}
\end{figure}

The results are presented in \cref{fig:mode-recall}.
With the \ac{MAP} mode alone, 29\% of signals are recovered; the configuration used for the examples (hypercube and symmetry searches with up to three modes) raises this to 38\%, and every search with up to fifteen modes to 50\%.
No search loses a signal recovered with the \ac{MAP} mode alone, and neither optional variant improves on the hypercube and symmetry searches.
Two conclusions follow.
First, the mode budget rather than the choice of search limits recovery: at five modes, adding the other searches to the hypercube search recovers one further signal, whereas raising the limit to fifteen modes recovers ten more.
We nonetheless use three modes in the examples (\cref{tab:settings}), since this study scores the seeding distribution rather than the corrected posteriors of \cref{sec:sampling}, and extra components are often mis-weighted: under Laplace weighting (\cref{sec:weights}), a secondary component outranks the \ac{MAP} mode in over a third of the mixtures in which both are present, mostly through its larger volume rather than a better fit.
Second, most of the remaining misses are due to the covariance rather than the search: in nearly all of the signals not recovered with fifteen modes, the injected parameters lie outside the standard deviation of a component with a higher log-posterior, consistent with a covariance that is too narrow rather than a missed mode.

\subsection{Optimisation of sequential Monte Carlo sampling}
\label{sec:settings}

\begin{figure*}
    \centering
    \includegraphics[width=0.49\textwidth]{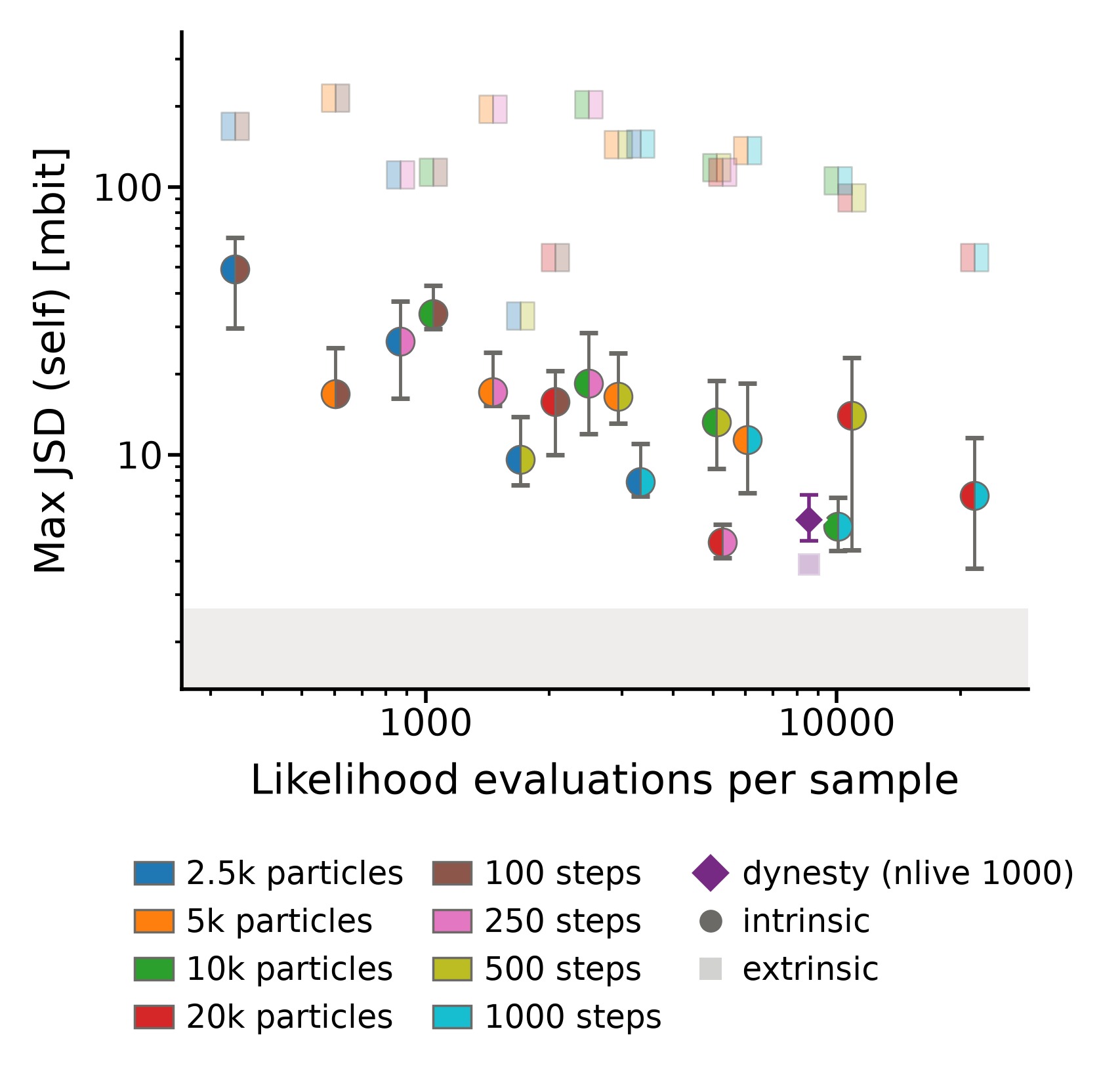}
    \hfill
    \includegraphics[width=0.49\textwidth]{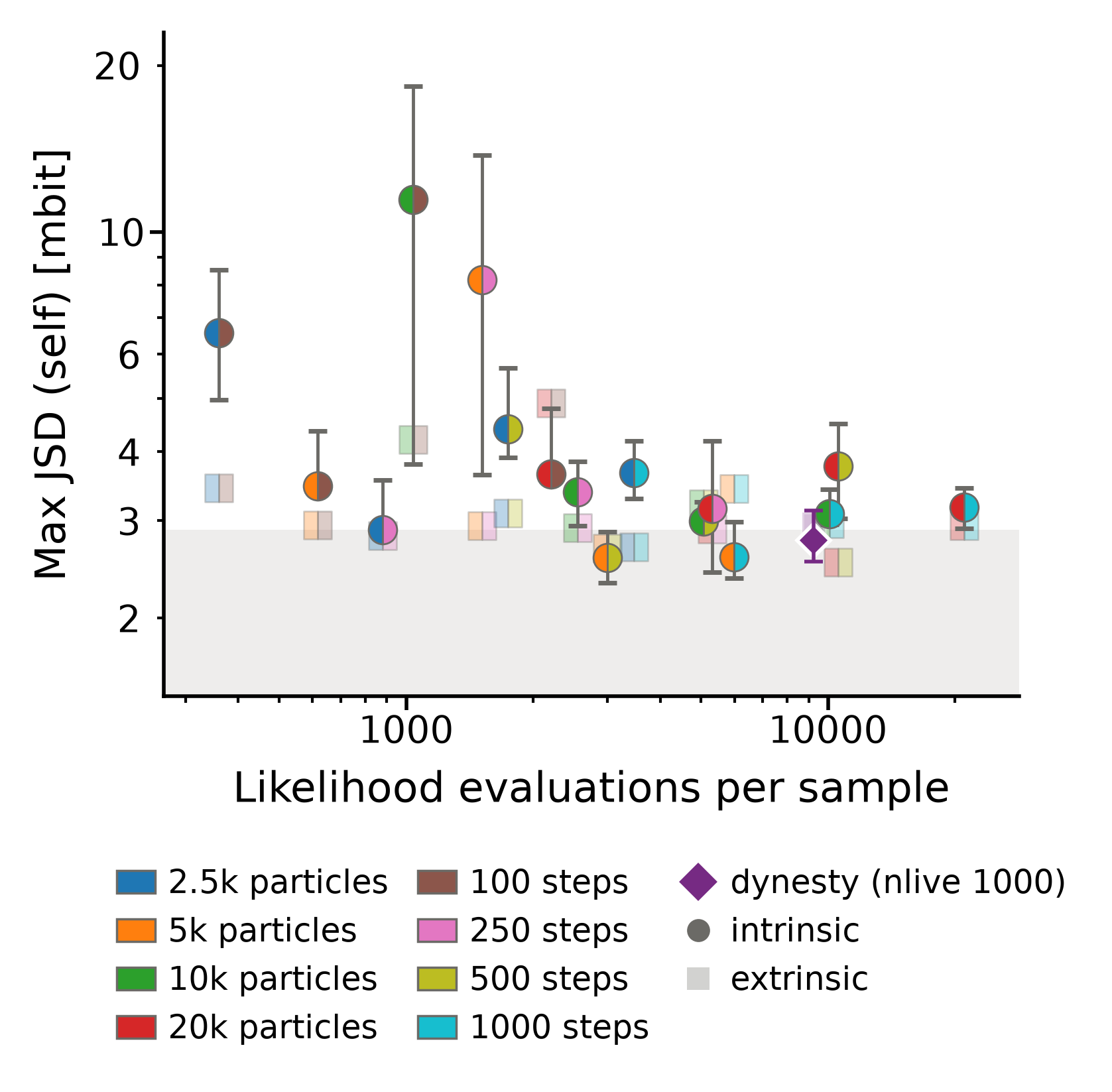}
    \caption{
    Run-to-run reproducibility of Laplace-seeded \aspire against cost, for the simulated precessing \ac{BBH} signal (BBH HLV; left) and the simulated \ac{BNS} signal (BNS 3G; right), over a grid of particle counts (left half of each marker) and \ac{MCMC} steps per iteration (right half).
    The vertical axis is the largest one-dimensional \ac{JSD} between the repeats of each configuration, and the horizontal axis the number of likelihood evaluations per posterior sample; \dynesty\ with 1000 live points is shown for comparison.
    Solid markers take the maximum over the intrinsic parameters, with error bars showing the 16--84\% interval from bootstrap resampling of the repeats, and translucent markers over the extrinsic parameters, where for BBH HLV the $\delta\phi$ mirror lobe dominates the divergence.
    The grey band marks the split-half noise floor.
    Note the narrower vertical range for the \ac{BNS}: every configuration lies within a factor of a few of the noise floor.
    }
    \label{fig:smc-settings}
\end{figure*}

Laplace-seeded \aspire combines the speed of seeding with the ability to correct an inaccurate seeding distribution, but its cost and reliability depend mainly on two settings: the number of particles, and the number of \ac{MCMC} steps used to move them at each iteration.
To choose these, we analysed two simulated signals: a precessing \ac{BBH} in simulated data from the \ac{LHO}, \ac{LLO} and Virgo network (BBH HLV, network \ac{SNR} 47), and a \ac{BNS} in simulated next-generation detector data (BNS 3G, \ac{SNR} 290).
Both are described, together with the other examples, in \cref{sec:simulated}, where they are analysed with the configuration chosen here (\cref{tab:summary}).
We ran each over a grid of $\{2500, 5000, 10000, 20000\}$ particles and $\{100, 250, 500, 1000\}$ steps, alongside \dynesty\ with 1000 live points.
We ran five repeats of each configuration, and ten of the configurations with 10000 particles and 250, 500 or 1000 steps.
Two of the ten random seeds failed to complete for BBH HLV in some configurations and were replaced with new seeds, which we then used for BNS 3G as well.
For the first, the seeding distribution included a third component from the hypercube search, and in two configurations the bisection for $\beta$ (\cref{sec:smc}) found no step larger than its tolerance after the first iteration; for the second, the tempering advanced so slowly that every configuration in which it ran reached its time limit, having taken 2.5--3.5 times as many iterations as another repeat.
To isolate the \ac{SMC} stage, the \ac{MAP} search is started from the injected parameters, so that every repeat begins from a seeding distribution built about the same \ac{MAP} point.
Since the true posterior is unknown, we measure reproducibility: the largest one-dimensional \ac{JSD} between the repeats of each configuration, a necessary but not sufficient condition for a configuration to be trusted.
Unlike the scored sets of \cref{sec:validation}, we take this maximum over all the sampled parameters, separately over the intrinsic parameters (masses, spins and tidal deformabilities) and the extrinsic parameters (the remainder), and estimate its uncertainty by bootstrap resampling of the repeats.

\Cref{fig:smc-settings} shows the results for BBH HLV, with the reproducibility of each configuration given separately for the intrinsic parameters (solid markers) and the extrinsic parameters (translucent markers).
As a benchmark, repeated \dynesty\ runs reproduce each other to 5.7\,mbit on the intrinsic parameters and 3.9\,mbit on the extrinsic parameters.
On the intrinsic parameters, the number of steps matters more than the number of particles.
With 100 steps, every configuration is poorly reproducible (16--49\,mbit), whatever the number of particles.
With more steps, four configurations reproduce the intrinsic parameters about as well as \dynesty\ does (4.7--7.9\,mbit), two of them at less than its cost per sample.
Between 7 and 20\,mbit, the uncertainties are too large to rank the configurations.
We adopt 10000 particles and 250 steps, which reaches 18\,mbit (16--84\% interval 12--29\,mbit) at 0.3 times the number of likelihood evaluations per sample of \dynesty; 20000 particles with 250 steps (4.7\,mbit) and 10000 with 1000 steps (5.4\,mbit) match \dynesty, but at roughly two and four times the cost.
For a Gaussian, 18\,mbit corresponds to a shift in the mean of ${\sim}0.3$ standard deviations, in the worst parameter between the most discrepant pair of repeats.
This exceeds the 7\,mbit threshold that \citet{LIGOScientific:2020ibl} used to flag differences between waveform models, but differences of this size arose for several of their events while remaining subdominant to the statistical uncertainty, and are comparable to those between waveform models and between samplers in recent catalogue analyses; we therefore prefer the cheaper configuration.
Even the most reproducible configurations differ systematically from \dynesty, by 7--15\,mbit, mostly in the spin magnitude $a_1$; since the \dynesty\ reference is under-walked on this problem (\cref{sec:dynesty-walks}), we cannot attribute this offset to either sampler.
We compare reproducibility and distance to \dynesty\ for every configuration in \cref{sec:bias-variance}.

On the extrinsic parameters (translucent markers), no configuration reproduces itself as well as \dynesty\ does (30--220\,mbit), and the largest divergence is always in $\delta\phi$.
The posterior has a second $\delta\phi$ lobe, offset by $\pi$ and carrying ${\sim}10\%$ of the posterior mass; the seeding distribution contains it, but the \ac{MCMC} moves cannot transfer particles between two narrow lobes separated by $\pi$, so individual runs lose or overweight it, and for our adopted configuration the ten repeats place between 0.7\% and 55\% of the mass there.
We report this as a limitation of the current implementation.

The right panel of \cref{fig:smc-settings} shows the same grid for BNS 3G, where the seeding distribution has a single mode in every repeat.
Here every configuration reproduces \dynesty\ on the extrinsic parameters (2.5--4.9\,mbit against 3.0\,mbit), and on the intrinsic parameters 13 of the 16 configurations lie at 2.6--4.4\,mbit against 2.8\,mbit; the other three each owe their divergence to a single outlying repeat with short \ac{MCMC} chains, and our adopted configuration reaches 3.4\,mbit.
Every configuration differs from \dynesty\ by a small systematic offset of 4--5.5\,mbit, mostly in $\chi_1$, as for $a_1$ on BBH HLV; in \cref{sec:bias-variance} we show that this offset appears even in configurations that reproduce themselves as well as \dynesty\ does, and that it more likely originates in the \dynesty\ reference than in \ac{SMC}.
Since the two grids share their code, seeds and configurations, the contrast between them localises the poor extrinsic reproducibility of BBH HLV to the $\delta\phi$ lobe rather than to the \ac{SMC} stage itself.

\section{Validation}
\label{sec:validation}
With the configuration fixed in \cref{sec:configuration}, we now present a series of validation studies to understand the performance and limitations of seeded sampling methods.
In \cref{tab:summary}, we summarise the metrics across all examples; the settings with which each method was run are given in \cref{tab:settings} of \cref{sec:run-settings}.
Specifically, we use the seeding distribution with the four sampling methods described in \cref{sec:sampling}, drawing in-prior samples with both the diagonal and the full truncation; not every method is run on every example.

Any such comparison has two axes, cost and accuracy, and every sampler trades one against the other through its settings, so the result depends on the configurations compared.
As the reference we use \dynesty\ with the standard \bilby\ configuration for gravitational-wave inference (\texttt{maxmcmc}${}=5000$), the baseline a user would otherwise run, and for the seeded methods we use the configuration chosen in \cref{sec:configuration}, held fixed across problems rather than tuned to each; better accuracy is available at a higher cost (\cref{sec:settings}).

We add two further analyses per example.
The first is \dynesty\ with the walk-length cap raised to 20\,000, since the standard cap binds on several of our problems (\cref{sec:dynesty-walks}).
The second is prior-initialised \aspire, which separates the effect of seeding from that of the sampler configuration: it differs from Laplace-seeded \aspire only in where the initial particles and the samples used to train the flow $q$ are drawn from, the prior rather than $g(\parameters)$, with no \ac{MAP} search or curvature computation, and with the same numbers of particles and \ac{MCMC} steps, schedule and mutation kernel (\cref{sec:smc,tab:settings}).
It is not a fair comparison of \aspire\ with \dynesty\ in its own right, since the configuration was chosen for the seeded analysis.

We report the cost of each method as its number of likelihood evaluations, in total and per independent posterior sample, and its speed-up as the ratio of the wall-clock sampling time of the \dynesty\ reference (\texttt{maxmcmc}${}=5000$) to that of the method.
We measure accuracy against \dynesty\ by the difference in log evidence, $\Delta\ln\evidence$, and by the one-dimensional \ac{JSD} between the marginal posteriors, given as its mean over parameters and its maximum.
For the compact-binary examples, we score a fixed set of physical parameters rather than every sampled coordinate, since several of the latter, such as the spin azimuthal angles, are barely constrained and dominate the divergence for reasons unrelated to the quantities of interest.
The intrinsic set is the detector-frame chirp mass and mass ratio, the effective inspiral spin $\chi_\mathrm{eff}$ and the effective precession spin $\chi_p$ (for the \ac{BNS}, whose spins are aligned, the tidal deformabilities $\Lambda_1$ and $\Lambda_2$ in place of $\chi_p$), and the extrinsic set is the inclination $\theta_{JN}$, the luminosity distance $d_L$ and the sky position.
We give the mean and maximum separately for each set.
The luminosity distance is marginalised in the likelihood, and so is reconstructed from the sampled parameters as in \cref{sec:pp}; the divergences for every sampled parameter are available in the data release.
Since two finite sets of samples from the same distribution do not give zero \ac{JSD}, we compare each value against a noise floor: the \ac{JSD} between disjoint subsets of the \dynesty\ samples, given on its row of \cref{tab:summary}.

\begin{table*}
    \centering
    \caption{
    Cost, speed-up, evidence difference and \ac{JSD} against the \dynesty\ reference (\texttt{maxmcmc}${}=5000$) for every method and example; noise floors are in brackets.
    Shading marks distance from the reference: \ac{JSD} is teal ($\leq 3$\,mbit) to yellow (50\,mbit), orange (50--100\,mbit) and dark red ($>100$\,mbit); speed-up is blue if faster and orange if slower; $|\Delta\ln\evidence|$ is orange.
    }
    \scriptsize
    \setlength{\tabcolsep}{4pt}
    \renewcommand{\arraystretch}{0.84}
\begin{tabular}{llrrrrrrlrl}
\toprule
 & & \multicolumn{2}{c}{$\nlikelihood$} & & & & \multicolumn{4}{c}{JSD [mbits]} \\
\cmidrule(lr){3-4} \cmidrule(lr){8-11}
Example & Method & Total & Per sample & Time & Speed-up & $\Delta\ln\evidence$ & \multicolumn{2}{c}{Mean} & \multicolumn{2}{c}{Max} \\
\midrule
Gaussian & Laplace & $6.2\times10^{2}$ & 0.062 & 6.0\,s & \cellcolor[RGB]{189,210,230}9.7$\times$ & $0.019$ & \multicolumn{2}{c}{\cellcolor[RGB]{111,184,180}3.0} & \multicolumn{2}{c}{\cellcolor[RGB]{111,184,180}$x$ (3.0)} \\
 & Laplace in-prior (diagonal) & $1.1\times10^{4}$ & 1.1 & 4.9\,s & \cellcolor[RGB]{183,206,228}12$\times$ & $0.019$ & \multicolumn{2}{c}{\cellcolor[RGB]{111,184,180}2.5} & \multicolumn{2}{c}{\cellcolor[RGB]{111,184,180}$y$ (2.5)} \\
 & Laplace in-prior (full) & $1.1\times10^{4}$ & 1.1 & 4.7\,s & \cellcolor[RGB]{182,205,228}12$\times$ & $0.019$ & \multicolumn{2}{c}{\cellcolor[RGB]{117,186,177}3.4} & \multicolumn{2}{c}{\cellcolor[RGB]{119,187,176}$y$ (3.5)} \\
 & Laplace + rejection & $5.5\times10^{4}$ & 5.5 & 5.1\,s & \cellcolor[RGB]{185,207,229}11$\times$ & $0.022 \pm 0.004$ & \multicolumn{2}{c}{\cellcolor[RGB]{117,186,177}3.4} & \multicolumn{2}{c}{\cellcolor[RGB]{124,189,174}$x$ (3.9)} \\
 & Laplace-seeded \aspire & $1.2\times10^{5}$ & 12 & 31\,s & \cellcolor[RGB]{237,242,248}1.9$\times$ & $0.01 \pm 0.01$ & \multicolumn{2}{c}{\cellcolor[RGB]{114,185,179}3.2} & \multicolumn{2}{c}{\cellcolor[RGB]{119,187,177}$y$ (3.5)} \\
 & Prior-initialised \aspire & $2.5\times10^{5}$ & 25 & 26\,s & \cellcolor[RGB]{231,239,246}2.3$\times$ & $0.03 \pm 0.03$ & \multicolumn{2}{c}{\cellcolor[RGB]{112,184,180}3.1} & \multicolumn{2}{c}{\cellcolor[RGB]{115,185,178}$y$ (3.2)} \\
 & \dynesty{} + PR & $1.0\times10^{5}$ & 50 & 22\,s & \cellcolor[RGB]{227,235,244}2.7$\times$ & $-0.02 \pm 0.04$ & \multicolumn{2}{c}{\cellcolor[RGB]{111,184,180}2.0} & \multicolumn{2}{c}{\cellcolor[RGB]{111,184,180}$y$ (2.1)} \\
 & \dynesty, \texttt{maxmcmc}=5k & $5.5\times10^{5}$ & 210 & 58\,s & 1.0$\times$ & $0 \pm 0.08$ & \multicolumn{2}{c}{(2.3)} & \multicolumn{2}{c}{--} \\
 & \dynesty, \texttt{maxmcmc}=20k & $5.5\times10^{5}$ & 210 & 59\,s & 0.99$\times$ & $0.00 \pm 0.08$ & \multicolumn{2}{c}{\cellcolor[RGB]{111,184,180}1.5} & \multicolumn{2}{c}{\cellcolor[RGB]{111,184,180}$x$ (1.6)} \\
\midrule
Rosenbrock & Laplace in-prior (diagonal) & $1.1\times10^{4}$ & 1.1 & 4.9\,s & \cellcolor[RGB]{177,201,226}15$\times$ & $-0.10$ & \multicolumn{2}{c}{\cellcolor[RGB]{241,234,119}39} & \multicolumn{2}{c}{\cellcolor[RGB]{253,239,114}$y$ (49)} \\
 & Laplace in-prior (full) & $1.1\times10^{4}$ & 1.1 & 4.8\,s & \cellcolor[RGB]{177,201,226}15$\times$ & $-0.10$ & \multicolumn{2}{c}{\cellcolor[RGB]{239,161,89}52} & \multicolumn{2}{c}{\cellcolor[RGB]{239,161,89}$y$ (52)} \\
 & Laplace + rejection & $4.0\times10^{5}$ & 40 & 9.5\,s & \cellcolor[RGB]{196,214,233}7.6$\times$ & $0.052 \pm 0.007$ & \multicolumn{2}{c}{\cellcolor[RGB]{111,184,180}2.6} & \multicolumn{2}{c}{\cellcolor[RGB]{114,185,178}$x$ (3.2)} \\
 & Laplace-seeded \aspire & $2.2\times10^{5}$ & 26 & 30\,s & \cellcolor[RGB]{230,238,246}2.4$\times$ & $0.01 \pm 0.02$ & \multicolumn{2}{c}{\cellcolor[RGB]{111,184,180}2.2} & \multicolumn{2}{c}{\cellcolor[RGB]{111,184,180}$x$ (2.6)} \\
 & Prior-initialised \aspire & $2.5\times10^{5}$ & 30 & 21\,s & \cellcolor[RGB]{219,230,242}3.5$\times$ & $0.04 \pm 0.03$ & \multicolumn{2}{c}{\cellcolor[RGB]{111,184,180}2.4} & \multicolumn{2}{c}{\cellcolor[RGB]{111,184,180}$x$ (2.7)} \\
 & \dynesty{} + PR & $7.0\times10^{5}$ & 310 & 1.8\,min & \cellcolor[RGB]{254,247,242}0.68$\times$ & $0.07 \pm 0.07$ & \multicolumn{2}{c}{\cellcolor[RGB]{111,184,180}1.2} & \multicolumn{2}{c}{\cellcolor[RGB]{111,184,180}$x$ (1.5)} \\
 & \dynesty, \texttt{maxmcmc}=5k & $7.5\times10^{5}$ & 310 & 1.2\,min & 1.0$\times$ & $0 \pm 0.07$ & \multicolumn{2}{c}{(1.6)} & \multicolumn{2}{c}{--} \\
 & \dynesty, \texttt{maxmcmc}=20k & $7.5\times10^{5}$ & 310 & 1.2\,min & 0.99$\times$ & $0.00 \pm 0.07$ & \multicolumn{2}{c}{\cellcolor[RGB]{111,184,180}0.93} & \multicolumn{2}{c}{\cellcolor[RGB]{111,184,180}$x$ (1.1)} \\
\midrule
AntiGlitch & Laplace in-prior (diagonal) & $1.7\times10^{4}$ & 1.7 & 12\,s & \cellcolor[RGB]{128,168,208}79$\times$ & $0.073$ & \multicolumn{2}{c}{\cellcolor[RGB]{111,184,180}2.7} & \multicolumn{2}{c}{\cellcolor[RGB]{111,184,180}$A$ (3.0)} \\
 & Laplace in-prior (full) & $1.7\times10^{4}$ & 1.7 & 14\,s & \cellcolor[RGB]{133,171,209}69$\times$ & $0.073$ & \multicolumn{2}{c}{\cellcolor[RGB]{111,184,180}2.8} & \multicolumn{2}{c}{\cellcolor[RGB]{112,184,180}$\log_{10}\gamma$ (3.1)} \\
 & Laplace + rejection & $2.5\times10^{5}$ & 24 & 2.2\,min & \cellcolor[RGB]{197,215,233}7.5$\times$ & $0.079 \pm 0.002$ & \multicolumn{2}{c}{\cellcolor[RGB]{111,184,180}2.3} & \multicolumn{2}{c}{\cellcolor[RGB]{111,184,180}$A$ (2.4)} \\
 & Laplace-seeded \aspire & $1.3\times10^{5}$ & 13 & 1.6\,min & \cellcolor[RGB]{187,208,230}10$\times$ & $0.09 \pm 0.01$ & \multicolumn{2}{c}{\cellcolor[RGB]{111,184,180}2.4} & \multicolumn{2}{c}{\cellcolor[RGB]{111,184,180}$A$ (2.8)} \\
 & Prior-initialised \aspire & $3.9\times10^{5}$ & 39 & 3.7\,min & \cellcolor[RGB]{212,225,239}4.5$\times$ & $0.13 \pm 0.04$ & \multicolumn{2}{c}{\cellcolor[RGB]{111,184,180}2.4} & \multicolumn{2}{c}{\cellcolor[RGB]{111,184,180}$\log_{10}\gamma$ (2.8)} \\
 & \dynesty{} + PR & $2.2\times10^{5}$ & 91 & 2.6\,min & \cellcolor[RGB]{201,218,235}6.4$\times$ & $0.09 \pm 0.05$ & \multicolumn{2}{c}{\cellcolor[RGB]{111,184,180}2.1} & \multicolumn{2}{c}{\cellcolor[RGB]{111,184,180}$A$ (2.4)} \\
 & \dynesty, \texttt{maxmcmc}=5k & $1.6\times10^{6}$ & 520 & 16\,min & 1.0$\times$ & $0 \pm 0.1$ & \multicolumn{2}{c}{(2.3)} & \multicolumn{2}{c}{--} \\
 & \dynesty, \texttt{maxmcmc}=20k & $1.6\times10^{6}$ & 520 & 17\,min & 0.97$\times$ & $0.0 \pm 0.1$ & \multicolumn{2}{c}{\cellcolor[RGB]{111,184,180}1.4} & \multicolumn{2}{c}{\cellcolor[RGB]{111,184,180}$f$ (1.6)} \\
\midrule
 & & & & & & & \multicolumn{2}{c}{Intrinsic} & \multicolumn{2}{c}{Extrinsic} \\
\cmidrule(lr){8-9} \cmidrule(lr){10-11}
 & & & & & & & Mean & Max & Mean & Max \\
\midrule
BNS 3G & Laplace in-prior (diagonal) & $4.5\times10^{4}$ & 4.5 & 41\,s & \cellcolor[RGB]{150,183,216}38$\times$ & \cellcolor[RGB]{244,187,145}$-3.7$ & \cellcolor[RGB]{196,99,103}260 & \cellcolor[RGB]{196,99,103}$\Lambda_1$ (380) & \cellcolor[RGB]{196,99,103}150 & \cellcolor[RGB]{196,99,103}$d_L$ (460) \\
SNR 290 & Laplace in-prior (full) & $4.5\times10^{4}$ & 4.5 & 41\,s & \cellcolor[RGB]{150,183,216}38$\times$ & \cellcolor[RGB]{244,187,145}$-3.7$ & \cellcolor[RGB]{239,161,89}94 & \cellcolor[RGB]{196,99,103}$\Lambda_2$ (170) & \cellcolor[RGB]{196,99,103}100 & \cellcolor[RGB]{196,99,103}$d_L$ (250) \\
 & Laplace-seeded \aspire & $2.5\times10^{7}$ & $2.5\times10^{3}$ & 8.9\,min & \cellcolor[RGB]{224,234,243}2.9$\times$ & \cellcolor[RGB]{253,244,237}$-0.46 \pm 0.03$ & \cellcolor[RGB]{111,184,180}3.0 & \cellcolor[RGB]{141,195,166}$\chi_\mathrm{eff}$ (5.5) & \cellcolor[RGB]{111,184,180}2.1 & \cellcolor[RGB]{111,184,180}$\alpha$ (2.6) \\
 & Prior-initialised \aspire & $5.3\times10^{7}$ & $5.3\times10^{3}$ & 14\,min & \cellcolor[RGB]{237,242,248}1.9$\times$ & \cellcolor[RGB]{254,247,243}$-0.40 \pm 0.04$ & \cellcolor[RGB]{111,184,180}2.7 & \cellcolor[RGB]{118,186,177}$\chi_\mathrm{eff}$ (3.5) & \cellcolor[RGB]{111,184,180}2.0 & \cellcolor[RGB]{111,184,180}$\delta$ (2.5) \\
 & \dynesty{} + PR & $2.8\times10^{7}$ & $6.1\times10^{3}$ & 20\,min & \cellcolor[RGB]{248,250,252}1.3$\times$ & \cellcolor[RGB]{252,237,225}$-0.6 \pm 0.2$ & \cellcolor[RGB]{111,184,180}2.5 & \cellcolor[RGB]{115,185,178}$\chi_\mathrm{eff}$ (3.2) & \cellcolor[RGB]{111,184,180}2.1 & \cellcolor[RGB]{111,184,180}$\delta$ (2.6) \\
 & \dynesty, \texttt{maxmcmc}=5k & $4.3\times10^{7}$ & $9.1\times10^{3}$ & 26\,min & 1.0$\times$ & $0 \pm 0.2$ & (2.1) & -- & (1.9) & -- \\
 & \dynesty, \texttt{maxmcmc}=20k & $8.0\times10^{7}$ & $1.7\times10^{4}$ & 54\,min & \cellcolor[RGB]{253,240,231}0.48$\times$ & \cellcolor[RGB]{254,250,247}$-0.4 \pm 0.2$ & \cellcolor[RGB]{111,184,180}2.2 & \cellcolor[RGB]{111,184,180}$\mathcal{M}$ (2.9) & \cellcolor[RGB]{111,184,180}2.2 & \cellcolor[RGB]{111,184,180}$\delta$ (2.9) \\
\midrule
BBH 3G & Laplace in-prior (diagonal) & $2.4\times10^{4}$ & 2.4 & 2.7\,min & \cellcolor[RGB]{122,163,205}180$\times$ & \cellcolor[RGB]{242,170,119}$-6.9$ & \cellcolor[RGB]{196,99,103}240 & \cellcolor[RGB]{196,99,103}$\chi_\mathrm{eff}$ (650) & \cellcolor[RGB]{196,99,103}180 & \cellcolor[RGB]{196,99,103}$\delta$ (260) \\
SNR 330 & Laplace in-prior (full) & $2.4\times10^{4}$ & 2.4 & 2.7\,min & \cellcolor[RGB]{122,163,205}180$\times$ & \cellcolor[RGB]{242,170,119}$-6.9$ & \cellcolor[RGB]{196,99,103}110 & \cellcolor[RGB]{196,99,103}$\chi_\mathrm{p}$ (200) & \cellcolor[RGB]{196,99,103}120 & \cellcolor[RGB]{196,99,103}$\delta$ (270) \\
 & Laplace-seeded \aspire & $3.5\times10^{7}$ & $3.5\times10^{3}$ & 2.1\,h & \cellcolor[RGB]{215,228,240}3.9$\times$ & \cellcolor[RGB]{247,202,170}$-2.11 \pm 0.03$ & \cellcolor[RGB]{151,199,161}6.7 & \cellcolor[RGB]{173,208,151}$\chi_\mathrm{eff}$ (10) & \cellcolor[RGB]{152,200,161}6.7 & \cellcolor[RGB]{167,206,154}$\theta_{JN}$ (9.1) \\
 & Prior-initialised \aspire & $6.3\times10^{7}$ & $6.3\times10^{3}$ & 3.4\,h & \cellcolor[RGB]{229,237,245}2.4$\times$ & \cellcolor[RGB]{246,198,163}$-2.47 \pm 0.04$ & \cellcolor[RGB]{199,218,139}17 & \cellcolor[RGB]{232,231,124}$q$ (32) & \cellcolor[RGB]{215,224,132}23 & \cellcolor[RGB]{242,235,119}$\theta_{JN}$ (39) \\
 & \dynesty{} + PR & $5.8\times10^{7}$ & $1.0\times10^{4}$ & 4.6\,h & \cellcolor[RGB]{238,243,249}1.8$\times$ & \cellcolor[RGB]{247,207,178}$-1.8 \pm 0.2$ & \cellcolor[RGB]{111,184,180}2.6 & \cellcolor[RGB]{111,184,180}$\mathcal{M}$ (2.8) & \cellcolor[RGB]{111,184,180}2.4 & \cellcolor[RGB]{111,184,180}$d_L$ (2.5) \\
 & \dynesty, \texttt{maxmcmc}=5k & $9.8\times10^{7}$ & $1.6\times10^{4}$ & 8.3\,h & 1.0$\times$ & $0 \pm 0.2$ & (2.3) & -- & (2.4) & -- \\
 & \dynesty, \texttt{maxmcmc}=20k & $3.6\times10^{8}$ & $6.0\times10^{4}$ & 34\,h & \cellcolor[RGB]{250,226,209}0.25$\times$ & \cellcolor[RGB]{251,230,215}$-0.7 \pm 0.2$ & \cellcolor[RGB]{111,184,180}2.7 & \cellcolor[RGB]{113,185,179}$\chi_\mathrm{p}$ (3.2) & \cellcolor[RGB]{111,184,180}2.4 & \cellcolor[RGB]{111,184,180}$\delta$ (2.6) \\
\midrule
BBH HLV & Laplace in-prior (diagonal) & $2.5\times10^{4}$ & 2.5 & 2.9\,min & \cellcolor[RGB]{127,167,207}83$\times$ & \cellcolor[RGB]{243,181,136}$-4.7$ & \cellcolor[RGB]{196,99,103}140 & \cellcolor[RGB]{196,99,103}$\chi_\mathrm{eff}$ (500) & \cellcolor[RGB]{196,99,103}200 & \cellcolor[RGB]{196,99,103}$d_L$ (520) \\
SNR 47 & Laplace in-prior (full) & $2.5\times10^{4}$ & 2.5 & 3.0\,min & \cellcolor[RGB]{128,167,207}81$\times$ & \cellcolor[RGB]{243,181,136}$-4.7$ & \cellcolor[RGB]{239,161,89}83 & \cellcolor[RGB]{196,99,103}$\chi_\mathrm{eff}$ (300) & \cellcolor[RGB]{196,99,103}130 & \cellcolor[RGB]{196,99,103}$d_L$ (340) \\
 & Laplace-seeded \aspire & $2.5\times10^{7}$ & $2.5\times10^{3}$ & 1.6\,h & \cellcolor[RGB]{229,237,245}2.5$\times$ & \cellcolor[RGB]{252,233,220}$-0.68 \pm 0.03$ & \cellcolor[RGB]{178,210,149}11 & \cellcolor[RGB]{208,222,135}$q$ (20) & \cellcolor[RGB]{155,201,159}7.2 & \cellcolor[RGB]{177,209,149}$\delta$ (11) \\
 & Prior-initialised \aspire & $4.0\times10^{7}$ & $4.0\times10^{3}$ & 2.2\,h & \cellcolor[RGB]{237,243,248}1.8$\times$ & \cellcolor[RGB]{249,218,196}$-1.17 \pm 0.03$ & \cellcolor[RGB]{150,199,162}6.5 & \cellcolor[RGB]{160,203,157}$\mathcal{M}$ (7.9) & \cellcolor[RGB]{125,189,174}3.9 & \cellcolor[RGB]{142,196,165}$\delta$ (5.6) \\
 & \dynesty{} + PR & $4.0\times10^{7}$ & $6.7\times10^{3}$ & 3.2\,h & \cellcolor[RGB]{248,250,252}1.3$\times$ & \cellcolor[RGB]{250,223,204}$-1.0 \pm 0.2$ & \cellcolor[RGB]{149,199,162}6.4 & \cellcolor[RGB]{163,204,156}$\chi_\mathrm{p}$ (8.4) & \cellcolor[RGB]{124,189,174}3.9 & \cellcolor[RGB]{160,203,157}$\delta$ (7.9) \\
 & \dynesty, \texttt{maxmcmc}=5k & $4.9\times10^{7}$ & $9.1\times10^{3}$ & 4.0\,h & 1.0$\times$ & $0 \pm 0.2$ & (2.1) & -- & (2.4) & -- \\
 & \dynesty, \texttt{maxmcmc}=20k & $1.3\times10^{8}$ & $2.4\times10^{4}$ & 12\,h & \cellcolor[RGB]{252,233,220}0.34$\times$ & \cellcolor[RGB]{253,243,236}$-0.5 \pm 0.2$ & \cellcolor[RGB]{123,188,175}3.8 & \cellcolor[RGB]{134,193,169}$\chi_\mathrm{eff}$ (4.8) & \cellcolor[RGB]{111,184,180}2.5 & \cellcolor[RGB]{115,185,178}$\delta$ (3.3) \\
\midrule
BBH2 HLV & Laplace in-prior (diagonal) & $4.2\times10^{4}$ & 4.2 & 5.6\,min & \cellcolor[RGB]{133,171,209}67$\times$ & \cellcolor[RGB]{240,162,105}$-9.4$ & \cellcolor[RGB]{196,99,103}110 & \cellcolor[RGB]{196,99,103}$\chi_\mathrm{eff}$ (250) & \cellcolor[RGB]{196,99,103}140 & \cellcolor[RGB]{196,99,103}$d_L$ (460) \\
SNR 40 & Laplace in-prior (full) & $4.2\times10^{4}$ & 4.2 & 5.6\,min & \cellcolor[RGB]{133,171,210}67$\times$ & \cellcolor[RGB]{240,162,105}$-9.4$ & \cellcolor[RGB]{196,99,103}110 & \cellcolor[RGB]{196,99,103}$\chi_\mathrm{eff}$ (230) & \cellcolor[RGB]{188,214,144}14 & \cellcolor[RGB]{220,226,129}$\theta_{JN}$ (26) \\
 & Laplace-seeded \aspire & $3.2\times10^{7}$ & $3.2\times10^{3}$ & 2.0\,h & \cellcolor[RGB]{222,232,243}3.1$\times$ & \cellcolor[RGB]{251,232,218}$-0.70 \pm 0.03$ & \cellcolor[RGB]{161,203,157}8.1 & \cellcolor[RGB]{204,220,137}$\chi_\mathrm{p}$ (19) & \cellcolor[RGB]{146,197,164}6.1 & \cellcolor[RGB]{173,208,151}$\theta_{JN}$ (10) \\
 & Prior-initialised \aspire & $4.5\times10^{7}$ & $4.5\times10^{3}$ & 2.3\,h & \cellcolor[RGB]{226,235,244}2.7$\times$ & $-0.33 \pm 0.03$ & \cellcolor[RGB]{111,184,180}2.9 & \cellcolor[RGB]{115,185,178}$\chi_\mathrm{p}$ (3.3) & \cellcolor[RGB]{111,184,180}3.0 & \cellcolor[RGB]{118,186,177}$\delta$ (3.4) \\
 & \dynesty{} + PR & $4.3\times10^{7}$ & $5.6\times10^{3}$ & 3.5\,h & \cellcolor[RGB]{238,243,249}1.8$\times$ & \cellcolor[RGB]{253,241,232}$-0.5 \pm 0.2$ & \cellcolor[RGB]{115,185,178}3.3 & \cellcolor[RGB]{126,189,173}$\mathcal{M}$ (4.0) & \cellcolor[RGB]{111,184,180}2.8 & \cellcolor[RGB]{124,189,174}$\delta$ (3.9) \\
 & \dynesty, \texttt{maxmcmc}=5k & $8.8\times10^{7}$ & $1.4\times10^{4}$ & 6.3\,h & 1.0$\times$ & $0 \pm 0.2$ & (2.4) & -- & (2.4) & -- \\
 & \dynesty, \texttt{maxmcmc}=20k & $1.1\times10^{8}$ & $1.6\times10^{4}$ & 7.6\,h & \cellcolor[RGB]{254,251,249}0.82$\times$ & $-0.2 \pm 0.2$ & \cellcolor[RGB]{111,184,180}2.8 & \cellcolor[RGB]{116,185,178}$q$ (3.3) & \cellcolor[RGB]{111,184,180}2.5 & \cellcolor[RGB]{111,184,180}$\alpha$ (2.8) \\
\midrule
GW150914 & Laplace in-prior (diagonal) & $2.8\times10^{5}$ & 28 & 5.6\,min & \cellcolor[RGB]{170,197,223}19$\times$ & \cellcolor[RGB]{241,164,109}$-8.6$ & \cellcolor[RGB]{196,99,103}120 & \cellcolor[RGB]{196,99,103}$\chi_\mathrm{eff}$ (270) & \cellcolor[RGB]{196,99,103}140 & \cellcolor[RGB]{196,99,103}$d_L$ (340) \\
SNR 24 & Laplace in-prior (full) & $2.8\times10^{5}$ & 28 & 5.6\,min & \cellcolor[RGB]{170,197,223}19$\times$ & \cellcolor[RGB]{241,164,109}$-8.6$ & \cellcolor[RGB]{239,161,89}60 & \cellcolor[RGB]{196,99,103}$q$ (140) & \cellcolor[RGB]{239,161,89}82 & \cellcolor[RGB]{196,99,103}$d_L$ (190) \\
 & Laplace-seeded \aspire & $1.9\times10^{7}$ & $1.9\times10^{3}$ & 1.2\,h & \cellcolor[RGB]{245,248,251}1.4$\times$ & $-0.09 \pm 0.02$ & \cellcolor[RGB]{111,184,180}2.5 & \cellcolor[RGB]{115,185,178}$\mathcal{M}$ (3.3) & \cellcolor[RGB]{130,191,171}4.4 & \cellcolor[RGB]{170,207,152}$\theta_{JN}$ (9.7) \\
 & Prior-initialised \aspire & $3.0\times10^{7}$ & $3.0\times10^{3}$ & 1.5\,h & \cellcolor[RGB]{250,252,253}1.2$\times$ & $-0.16 \pm 0.03$ & \cellcolor[RGB]{111,184,180}3.0 & \cellcolor[RGB]{129,191,171}$\mathcal{M}$ (4.3) & \cellcolor[RGB]{124,189,174}3.9 & \cellcolor[RGB]{147,198,163}$\theta_{JN}$ (6.2) \\
 & \dynesty{} + PR & $2.2\times10^{7}$ & $4.5\times10^{3}$ & 1.7\,h & 1.0$\times$ & $-0.1 \pm 0.1$ & \cellcolor[RGB]{111,184,180}2.4 & \cellcolor[RGB]{121,187,175}$\mathcal{M}$ (3.7) & \cellcolor[RGB]{111,184,180}2.7 & \cellcolor[RGB]{127,190,173}$\theta_{JN}$ (4.1) \\
 & \dynesty, \texttt{maxmcmc}=5k & $2.4\times10^{7}$ & $4.7\times10^{3}$ & 1.7\,h & 1.0$\times$ & $0 \pm 0.2$ & (2.3) & -- & (2.4) & -- \\
 & \dynesty, \texttt{maxmcmc}=20k & $6.8\times10^{7}$ & $1.3\times10^{4}$ & 5.4\,h & \cellcolor[RGB]{251,232,217}0.32$\times$ & $0.2 \pm 0.2$ & \cellcolor[RGB]{111,184,180}2.3 & \cellcolor[RGB]{116,186,178}$\mathcal{M}$ (3.3) & \cellcolor[RGB]{111,184,180}2.4 & \cellcolor[RGB]{111,184,180}$\delta$ (3.0) \\
\midrule
GW240615 & Laplace in-prior (diagonal) & $4.7\times10^{5}$ & 47 & 6.6\,min & \cellcolor[RGB]{168,195,222}20$\times$ & \cellcolor[RGB]{240,161,104}$-9.8$ & \cellcolor[RGB]{196,99,103}320 & \cellcolor[RGB]{196,99,103}$q$ (610) & \cellcolor[RGB]{253,239,114}49 & \cellcolor[RGB]{196,99,103}$d_L$ (140) \\
SNR 23 & Laplace in-prior (full) & $4.7\times10^{5}$ & 47 & 6.7\,min & \cellcolor[RGB]{168,195,223}20$\times$ & \cellcolor[RGB]{240,161,104}$-9.8$ & \cellcolor[RGB]{196,99,103}390 & \cellcolor[RGB]{196,99,103}$q$ (650) & \cellcolor[RGB]{243,235,118}41 & \cellcolor[RGB]{196,99,103}$d_L$ (140) \\
 & Laplace-seeded \aspire & $2.7\times10^{7}$ & $2.7\times10^{3}$ & 1.7\,h & \cellcolor[RGB]{247,249,252}1.3$\times$ & \cellcolor[RGB]{250,220,199}$-1.08 \pm 0.02$ & \cellcolor[RGB]{237,233,121}36 & \cellcolor[RGB]{239,161,89}$q$ (74) & \cellcolor[RGB]{202,219,137}18 & \cellcolor[RGB]{251,238,114}$\theta_{JN}$ (48) \\
 & Prior-initialised \aspire & $3.3\times10^{7}$ & $3.3\times10^{3}$ & 1.6\,h & \cellcolor[RGB]{245,248,251}1.4$\times$ & \cellcolor[RGB]{253,245,239}$-0.43 \pm 0.03$ & \cellcolor[RGB]{118,186,177}3.4 & \cellcolor[RGB]{134,193,169}$\mathcal{M}$ (4.8) & \cellcolor[RGB]{191,215,143}14 & \cellcolor[RGB]{235,232,122}$\delta$ (34) \\
 & \dynesty{} + PR & $4.6\times10^{7}$ & $4.7\times10^{3}$ & 3.9\,h & \cellcolor[RGB]{253,244,237}0.58$\times$ & $-0.3 \pm 0.2$ & \cellcolor[RGB]{187,213,144}14 & \cellcolor[RGB]{235,232,122}$\mathcal{M}$ (34) & \cellcolor[RGB]{144,196,165}5.8 & \cellcolor[RGB]{171,207,152}$d_L$ (9.8) \\
 & \dynesty, \texttt{maxmcmc}=5k & $3.2\times10^{7}$ & $5.2\times10^{3}$ & 2.2\,h & 1.0$\times$ & $0 \pm 0.2$ & (2.2) & -- & (2.5) & -- \\
 & \dynesty, \texttt{maxmcmc}=20k & $8.1\times10^{7}$ & $1.3\times10^{4}$ & 6.5\,h & \cellcolor[RGB]{252,233,220}0.35$\times$ & $0.0 \pm 0.2$ & \cellcolor[RGB]{111,184,180}2.3 & \cellcolor[RGB]{111,184,180}$\chi_\mathrm{eff}$ (2.6) & \cellcolor[RGB]{111,184,180}2.5 & \cellcolor[RGB]{112,184,180}$\delta$ (3.1) \\
\bottomrule
\end{tabular}

    \label{tab:summary}
\end{table*}

\subsection{Low-dimensional examples}
\label{sec:analytic}

\begin{figure}
    \centering
    \includegraphics[width=\columnwidth]{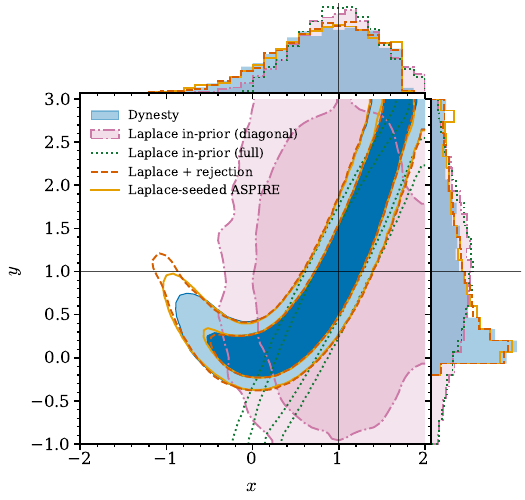}
    \caption{
    Posterior for the two-dimensional Rosenbrock likelihood from \dynesty (filled) and from the seeding distribution, drawn directly with independent (diagonal) and with full covariance, and corrected by rejection sampling and by Laplace-seeded \aspire.
    Contours enclose 90\% and 99\% of the probability; the black lines mark the maximum at $(1, 1)$.
    }
    \label{fig:rosenbrock}
\end{figure}

We begin with three low-dimensional problems: one where the approach works optimally (a correlated Gaussian likelihood, for which the Laplace approximation is exact), one where it fails by design (the curved Rosenbrock likelihood), and one simple gravitational-wave problem (a simulated non-Gaussian transient).

For the first case, we use a two-dimensional correlated Gaussian likelihood, with standard deviations 0.3 and 0.5 and correlation 0.7, under a uniform prior much wider than the posterior.
The results across a range of methods are presented in \cref{tab:summary}.
Since these problems are cheap and low-dimensional, the Laplace-seeded and prior-initialised \aspire\ runs use a smaller configuration than the gravitational-wave examples, with 5000 particles and 5 \ac{MCMC} steps per iteration (\cref{tab:settings}), which we justify at the end of this section.
Here the Laplace approximation is exact, so this tests only the implementation: the raw Laplace approximation agrees with \dynesty to within ${\sim}1$\,mbit of the noise floor of the \ac{JSD} (3.0\,mbit against 2.3\,mbit), and every method recovers the evidence.
Drawing from the seeding distribution costs 1.1 likelihood evaluations per sample, against 210 for \dynesty.

Next is the two-dimensional Rosenbrock likelihood, $\ln\likelihood = -(1-x)^2 - 100(y-x^2)^2$, whose curved, banana-shaped posterior is far from Gaussian and a standard test of sampling methods.
At the maximum $(1,1)$, the Laplace covariance has correlation 0.999, a thin sliver along the tangent to the banana.
Moreover, the prior bounds truncate the posterior support.
In \cref{fig:rosenbrock}, we show the posterior approximated by the in-prior diagonal and full Laplace covariances, the posterior recovered by \dynesty (from the full prior), and the posteriors recovered from the seeding distribution by rejection sampling and by Laplace-seeded \aspire initialised from the in-prior diagonal seeding distribution.
Both the in-prior diagonal and full covariances are poor approximations, with \ac{JSD} 39--52\,mbit against \dynesty, while rejection sampling and Laplace-seeded \aspire both recover the posterior close to the noise floor of the \ac{JSD}.
In \cref{tab:summary}, we report the full metrics from the full set of methods.
These further reveal the primary failure mode of our approach: when the posterior is far from Gaussian, the Laplace approximation is poor and the seeding distribution is a poor match to the posterior, so that seeding gains little.
Laplace-seeded \aspire is slower than prior-initialised \aspire (speed-ups of 2.4 and 3.5), and posterior repartitioning is slower than \dynesty (speed-up 0.68).
Rejection sampling is the fastest accurate method here (speed-up 7.6).

Finally, we consider an application beyond astrophysical signals: \citet{Ashton:2026seh} measure the rate of non-Gaussian noise artefacts (known commonly as glitches \citep{Davis:2022dnd}) in gravitational-wave detectors with a hierarchical Bayesian model, which requires parameter estimation with the \texttt{antiglitch} model \citep{Bondarescu:2023jcx} for a large population of glitches, so that the cost of each analysis matters.
We simulate a glitch with this model in a single LIGO detector at the high end of the projected O4 sensitivity \citep{KAGRA:2020npa}.
The \texttt{antiglitch} model is a frequency-domain model characterised by an amplitude $A$, a Gaussian envelope in log-frequency with central frequency $f$ and width set by $\gamma$, a phase and a time.
We inject the glitch with \ac{SNR} 20 and marginalise over the phase and time, leaving three sampled parameters.
This posterior is close to Gaussian, and every method agrees with \dynesty\ at the noise floor, with speed-ups over \dynesty from ${\sim}4.5$ for prior-initialised \aspire to 70--80 for Laplace in-prior sampling.
For population studies of this kind, the Laplace approximation could therefore be used directly, without correction, to accelerate the analysis of each glitch.

We chose the smaller \ac{SMC} configuration for these three problems from a separate grid study of 1000--10\,000 particles and 5--250 steps, with three repeats per cell.
Every cell reproduced \dynesty\ at the noise floor on all three problems, except for 1000 particles with 5 steps on the Rosenbrock problem, where one of three repeats was poor; we therefore adopt 5000 particles, a factor of five above that cell, with 5 steps.

\subsection{Simulated compact binary signals}
\label{sec:simulated}
We now turn to four simulated compact binary signals, each injected into Gaussian noise coloured by the detector \acp{PSD}, for which the injected parameters are known.
We summarise their properties and the settings used to analyse them in \cref{tab:settings}.
The first is a \ac{BNS} signal, labelled BNS 3G in \cref{tab:summary}, observed by a network of Cosmic Explorer at the \ac{LHO} site \citep{Evans:2016mbw}, LIGO-India \citep{M1100296} at A+ sensitivity \citep{T1800042, KAGRA:2020npa},\footnote{
    Specifically, we use the A+ amplitude spectral density distributed with \bilby~2.8.2 as \texttt{Aplus\_asd.txt}, the default for its \texttt{A1} detector.
    This is sampled at the same frequencies as the curve published in Ref.~\cite{T1800042}, but differs from it by up to 13\%, being more sensitive between 40 and 300\,Hz and less sensitive above 1\,kHz.
} and \ac{LLO} at the high end of the projected O4 sensitivity \citep{KAGRA:2020npa}, with a network \ac{SNR} of 290.
We analyse it with the aligned-spin, tidal \texttt{IMRPhenomD\_NRTidalv2} waveform model \citep{Husa:2015iqa, Khan:2015jqa, Dietrich:2019kaq} and a relative-binning likelihood \citep{Zackay:2018qdy, Cornish:2021lje, Krishna:2023bug}, marginalising over distance and phase, leaving 11 sampled parameters.
This is the regime in which we expect the Laplace approximation to be most useful: the signal is loud, and the posterior is correspondingly narrow compared to the prior.
The remaining three are precessing \acp{BBH}, analysed with \texttt{IMRPhenomXPHM} and marginalising over distance and time, leaving 13 sampled parameters.
BBH 3G and BBH HLV are the same near face-on, near equal-mass system ($q=0.8$, $\theta_{JN}=0.4$), observed in the next-generation network of BNS 3G with network \ac{SNR} 330, almost all of it in Cosmic Explorer, and in the \ac{LHO}, \ac{LLO} and Virgo network with network \ac{SNR} 47; the waveform, frequency band, priors and sampler settings are otherwise identical, so that the two differ only in the network.
BBH2 HLV is chosen to favour precession and higher-order modes: it is more asymmetric ($q=0.5$), more inclined ($\theta_{JN}=1.0$), and has larger in-plane spins, with network \ac{SNR} 40 in the \ac{LHO}, \ac{LLO} and Virgo network.
In all four cases, the \ac{MAP} search is started from the injected parameters: this is representative of what may be done in a forecasting study of simulated signals.

\begin{figure*}
    \centering
    \includegraphics[width=0.49\textwidth]{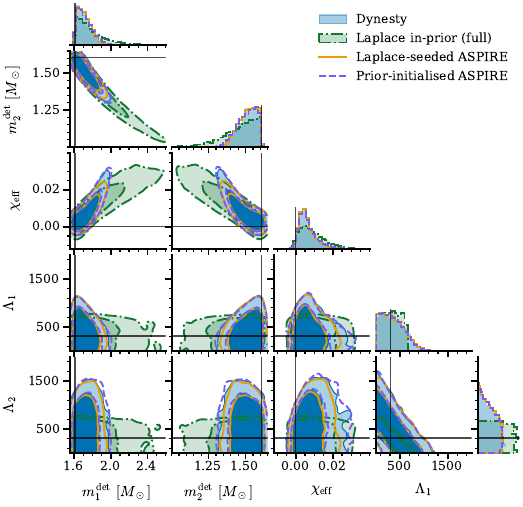}
    \hfill
    \includegraphics[width=0.49\textwidth]{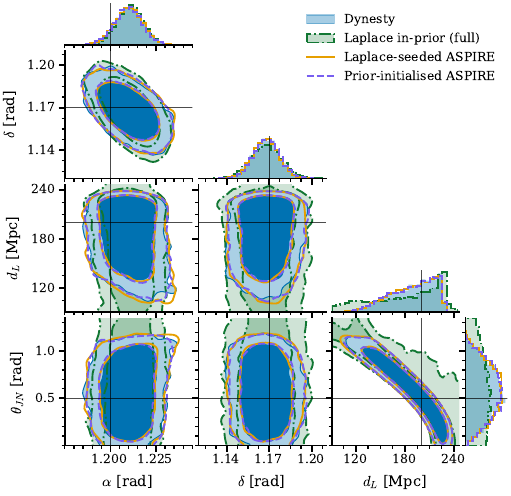}
    \caption{
    Posterior for the simulated \ac{BNS} signal in a next-generation network (BNS 3G; \cref{tab:summary}) from \dynesty (filled), Laplace in-prior (full), Laplace-seeded \aspire, and prior-initialised \aspire.
    Left: detector-frame component masses, effective inspiral spin and tidal deformabilities; right: sky position, luminosity distance and inclination.
    Contours enclose 90\% and 99\% of the probability, and black lines mark the injected values.
    }
    \label{fig:bns-corner}
\end{figure*}

\Cref{fig:bns-corner} compares the posteriors for the \ac{BNS} from \dynesty, Laplace in-prior sampling with the full covariance, and Laplace-seeded and prior-initialised \aspire.
Laplace in-prior sampling costs a few likelihood evaluations per sample, but the result, though a reasonable first look, is not statistically consistent with the posterior: the largest divergences are in the tidal deformabilities, whose posteriors are bounded by the prior at zero and far from Gaussian, and in the distance.
Using the full rather than the diagonal covariance reduces the divergence in both, but does not close the gap.
Laplace-seeded \aspire\ corrects this, agreeing with \dynesty\ close to the noise floor on every parameter at a wall-clock speed-up of 2.9.
Prior-initialised \aspire\ and posterior repartitioning reach the same accuracy at smaller speed-ups of 1.9 and 1.3.

\begin{figure*}
    \centering
    \includegraphics[width=0.49\textwidth]{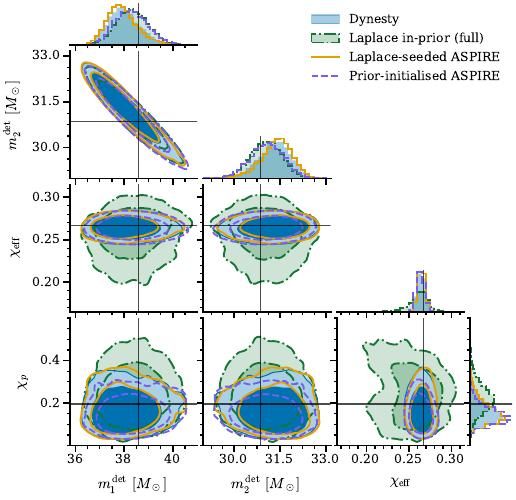}
    \hfill
    \includegraphics[width=0.49\textwidth]{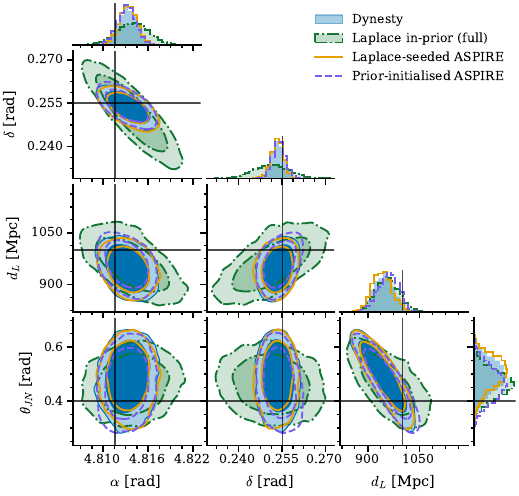}
    \caption{
    As \cref{fig:bns-corner}, for the simulated precessing \ac{BBH} signal in the next-generation network (BBH 3G).
    Left: detector-frame component masses, effective inspiral spin and effective precession spin; right: sky position, luminosity distance and inclination.
    The Laplace approximation is close to the posterior in the masses but several times too wide in $\chi_\mathrm{eff}$ and two to three times too wide in the sky position.
    }
    \label{fig:bbh3g-corner}
\end{figure*}

\Cref{fig:bbh3g-corner} compares the posteriors for BBH 3G from the same methods.
The posterior has a single mode in every method, \dynesty included, as does the seeding distribution.
The in-prior draws are again cheap but not statistically consistent with \dynesty: the Laplace approximation is close to the posterior in the component masses, but several times too wide in $\chi_\mathrm{eff}$ and two to three times too wide in the sky position.
Posterior repartitioning agrees with \dynesty\ at the noise floor at a speed-up of 1.8.
Laplace-seeded \aspire\ is the fastest corrected method, at a speed-up of 3.9, but not quite as accurate: its mean \acp{JSD} of ${\sim}7$\,mbit, largest in $\chi_\mathrm{eff}$ and $\theta_{JN}$, are visible as a subtle shift in the component masses and effective spins.
Prior-initialised \aspire\ is both slower (speed-up 2.4) and less accurate, with mean \acp{JSD} of 17--23\,mbit.
This run is not stable from run to run: a repeat with identical settings gave a mean \ac{JSD} of ${\sim}250$\,mbit and an evidence lower by ${\sim}5$, so the result above is a single draw from a sampler that can also fail on this example; we do not explore this run-to-run variation further, as it is beyond the scope of this work.

For the two \acp{BBH} in the \ac{LHO}, \ac{LLO} and Virgo network, in-prior draws are again cheap but not statistically consistent with \dynesty, with the full and diagonal covariances differing from it by similar amounts.
The worst-recovered intrinsic parameter is $\chi_\mathrm{eff}$, as for BBH 3G, and in the diagonal case the worst extrinsic parameter is $d_L$, which reflects the bias of in-prior draws that fit the data poorly (\cref{sec:pp}).
Posterior repartitioning recovers both posteriors close to the noise floor at a speed-up of 1.3--1.8.
Laplace-seeded \aspire\ is faster (2.5--3.1) but less accurate, with mean intrinsic \acp{JSD} of 8--11\,mbit, and prior-initialised \aspire\ with the same settings is more accurate than either on both signals.
On BBH HLV, the largest divergence is in the mass ratio, where Laplace-seeded \aspire overweights a region near equal mass and large $\theta_1$ by roughly a factor of seven relative to \dynesty.
This region is close to an additional component, placed at the prior boundary $q=1$ by the hypercube search, and the excess grows during the final tempering steps rather than being inherited from the seeding distribution.
On BBH2 HLV, the largest divergence is in $\chi_p$.
We regard these residual differences of at most 20\,mbit, like those on BBH 3G, as low-level sampler systematics rather than a failure of the seeding: the dependence of the reproducibility of \ac{SMC} on its settings in \cref{sec:settings} suggests that they could likely be reduced by increasing the particles or \ac{MCMC} steps, at a proportionate cost.

Finally, the evidence estimates of the seeded methods are systematically lower than that of the \dynesty reference on the simulated signals and GW240615, by 0.3--1.2 for Laplace-seeded and prior-initialised \aspire and posterior repartitioning, and by 1.8--2.5 on BBH 3G, which is larger than the quoted uncertainties; on GW150914 the offset is at most 0.2.
However, \dynesty with the raised \texttt{maxmcmc} cap also moves in the same direction on the simulated signals, by 0.2--0.7, suggesting that part of this offset is due to the reference rather than the seeded methods; we discuss this further in \cref{sec:dynesty-walks}.

\subsection{Gravitational-wave events}
\label{sec:events}

\begin{figure*}
    \centering
    \includegraphics[width=0.49\textwidth]{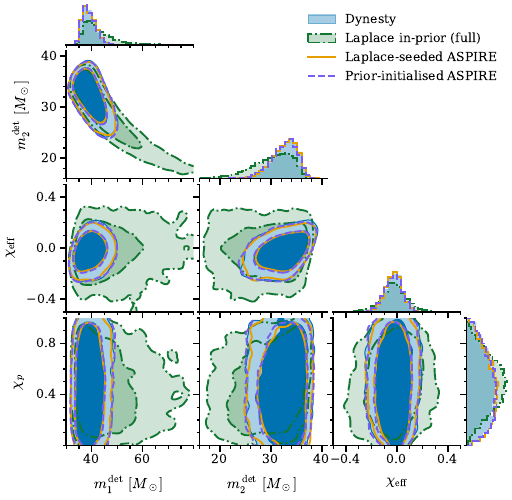}
    \hfill
    \includegraphics[width=0.49\textwidth]{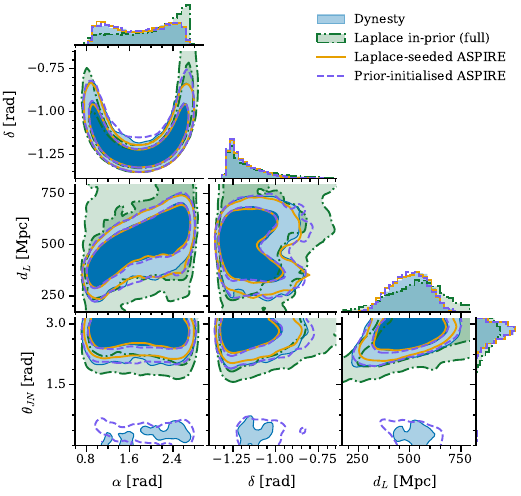}
    \caption{
    As \cref{fig:bns-corner}, for GW150914, for which there are no injected values.
    Left: detector-frame component masses, effective inspiral spin and effective precession spin; right: sky position, luminosity distance and inclination.
    The small face-on mode at $\theta_{JN} \lesssim 0.8$\,rad is recovered by \dynesty and prior-initialised \aspire, but is nearly absent from the Laplace-seeded \aspire posterior.
    }
    \label{fig:gw150914-corner}
\end{figure*}

Finally, we analyse two real signals: GW150914 \citep{LIGOScientific:2016aoc}, observed by \ac{LHO} and \ac{LLO} \citep{TheLIGOScientific:2016agk}, and GW240615\_113620 \citep{LIGOScientific:2026wfs} (hereafter GW240615), observed by \ac{LHO}, \ac{LLO} \citep{Capote:2024rmo, LIGO:2024kkz, LIGO:2026ika} and Virgo \citep{VIRGO:2026okv}, with network \acp{SNR} of 24 and 23 respectively.
For each, we analyse 4\,s of open data \citep{LIGOScientific:2026jgl} between 20 and 512\,Hz, with the \ac{PSD} estimated by a median average over 128\,s of data preceding the analysis segment.
We use \texttt{IMRPhenomXPHM} with spin-precession angles from integrating the SpinTaylor equations \citep{Pratten:2020ceb, Colleoni:2024knd}, with the remaining configuration given in \cref{tab:settings}; as for the simulated \acp{BBH}, we marginalise over distance and time, leaving 13 sampled parameters.
Unlike the simulated signals, the \ac{MAP} search is now blind (\cref{tab:settings}), and this dominates the cost of Laplace in-prior sampling, which is ten times that of the simulated \acp{BBH} per sample, though still a factor of ${\sim}20$ faster than \dynesty.
As for the simulated \acp{BBH}, in-prior draws are a reasonable first look but not statistically consistent with \dynesty, with the largest divergences mostly in the mass ratio and $\chi_\mathrm{eff}$, and far larger for GW240615 than for GW150914.

\Cref{fig:gw150914-corner} compares the posteriors for GW150914 from the same methods.
All three corrected methods agree with \dynesty\ close to the noise floor at speed-ups of 1.0--1.4, and their evidences agree to within 0.3.
The methods are indistinguishable in the main peak of the posterior, and differ instead in the weight they assign to a small secondary mode at face-on inclination ($\theta_{JN} \lesssim 0.8$\,rad).
\dynesty places 1.3\% of its samples in this mode and prior-initialised \aspire 2.5\%, while Laplace-seeded \aspire places almost none.
The mode lies far from the \ac{MAP} point, and none of the secondary-mode searches of \cref{sec:modes} admits a component there, so the seeding distribution places essentially no particles in it, and the \ac{MCMC} moves of \ac{SMC} rarely carry particles across to it, as for the $\delta\phi$ lobes of \cref{sec:settings}.
Because the mode carries only ${\sim}2\%$ of the posterior mass, it barely affects the mean \ac{JSD}, and so the summary metrics in \cref{tab:summary} understate this difference.
Drawing a fraction of the initial cloud from the prior (\cref{sec:smc}) is one possible remedy, since it places particles in regions the mode search did not reach, but in our tests it did not reliably work: at 300 steps a fraction of 0.1 recovered 3.0\% of the samples in the face-on mode on one seed, while at 250 steps a fraction of 0.2 overweighted it (9.5\%).
This shows that, for current-generation events with significant secondary modes, seeded runs are not yet the right tool without further work to make the identification of multiple modes more robust.

For GW240615, prior-initialised \aspire\ agrees with \dynesty\ in the intrinsic parameters at a speed-up of 1.4, though less well in the extrinsic parameters, where the largest divergence is in the declination; the seeded methods are less successful.
Posterior repartitioning reaches 14\,mbit in the chirp mass and is slower than \dynesty\ (speed-up 0.58), and Laplace-seeded \aspire\ reaches 36\,mbit in the mass ratio at a speed-up of 1.3, with an evidence lower than \dynesty's by ${\sim}1$.

\begin{figure}
    \centering
    \includegraphics[width=\columnwidth]{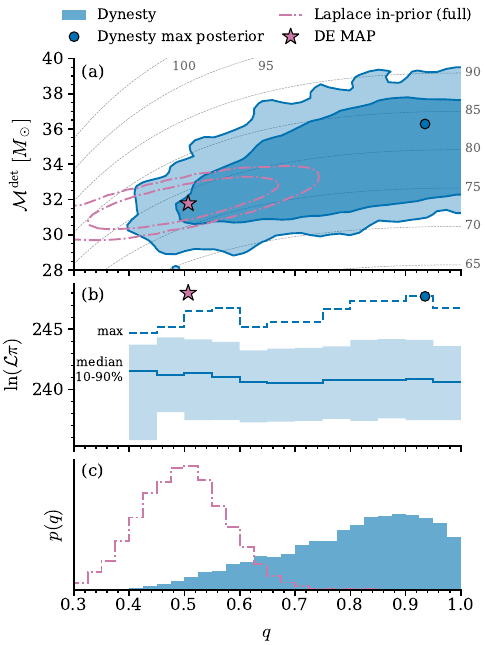}
    \caption{
    The posterior of GW240615 along the mass-ratio ridge.
    (a) Detector-frame chirp mass against mass ratio for \dynesty (filled) and the full-covariance Laplace in-prior draw (dash-dot), with 90\% and 99\% contours; grey curves are lines of constant detector-frame total mass, labelled in $M_\odot$.
    The star marks the \ac{MAP} and the circle the highest-posterior \dynesty sample.
    (b) The log posterior density of \dynesty binned in $q$: median and 10--90\% band, and the highest sample in each bin (dashed).
    (c) The mass-ratio marginals.
    }
    \label{fig:gw240615-ridge}
\end{figure}

\Cref{fig:gw240615-ridge} shows why the in-prior draw fails on GW240615, and that this is not a failed \ac{MAP} search.
The event has a single mode, and the \ac{MAP} is real, but it lies at one end of a long ridge along which the posterior density is nearly flat.
The detector-frame total mass is measured, but at a network \ac{SNR} of 23 the mass ratio is not, so the posterior extends along a line of constant total mass, with the chirp mass rising with the mass ratio.
Binned in $q$, the median log posterior density of \dynesty varies by less than 1 across $q = 0.4$--$1$, and the highest-posterior \dynesty sample (at $q \approx 0.94$) is only 0.27 below the \ac{MAP} (at $q \approx 0.51$).
The posterior mass therefore follows prior volume, with 77\% at $q > 0.7$, whereas the Gaussian built at the \ac{MAP} has the narrow width of that end of the ridge and barely reaches $q = 0.65$.
Laplace-seeded \aspire starts from this proposal and must move along the ridge, which it does only partly in the 250 steps used (median $q$ of 0.71 against 0.82 for \dynesty).
Posterior repartitioning has the same difficulty, since its effective prior is this narrow distribution: the nested sampler must expand along the ridge from live points that start at its low-$q$ end, which costs the speed-up (0.58) and leaves a residual divergence in the chirp mass, the other coordinate of the ridge.
GW150914, at a similar \ac{SNR}, has the same ridge, but a shorter one with its \ac{MAP} in the flat bulk rather than at one end, so that its in-prior draw covers too much of $q$ rather than too little, which is the safe direction for the corrected methods (\cref{sec:gw150914-ridge}).

\subsection{Prior volume}
\label{sec:prior-volume}
We expect a practical advantage of model-informed seeding to be that its cost depends only weakly on the width of the prior: the \ac{MAP} search locates the posterior by optimisation, and the samplers then start from the seeding distribution rather than from the prior.
By contrast, nested sampling must contract its live points from the full prior down to the posterior, and \ac{SMC} initialised from the prior must temper across it, so the cost of both grows with the prior volume.
To test this, we take the AntiGlitch example of \cref{sec:analytic} and widen the prior on the amplitude $A$ from $[0, 10^2]$, the prior used in \cref{tab:summary}, to $[0, 10^5]$ over seven widths.
Since the posterior on $A$ has a width of 1.0 (the standard deviation of the \dynesty posterior at the narrowest prior), and we confirm below that it does not change with the prior width, this varies the ratio of the prior width to the posterior width from $10^2$ to $10^5$ while leaving the posterior itself unchanged.
Unlike in \cref{sec:analytic}, the \ac{MAP} search is not started from the injected parameters, so the cost of \ac{DE} is included at every width.
We run each method three times at each width, on a single core; the \ac{SMC} runs use the low-dimensional configuration of \cref{sec:analytic}.
For both \aspire runs, we reduce the tolerance of the bisection for $\beta$ (\cref{sec:smc}) from $10^{-6}$ to $10^{-12}$: the first tempering step of prior-initialised \aspire falls as the square of the prior width, from ${\sim}2\times10^{-3}$ at the narrowest prior to ${\sim}10^{-9}$ at the widest, so at the default tolerance it cannot take a first step for ratios of $10^{4}$ and above.
The Laplace-seeded runs are identical at either tolerance.

\begin{figure}
    \centering
    \includegraphics[width=\columnwidth]{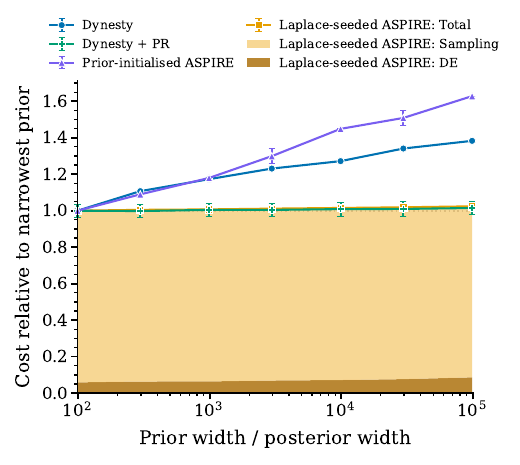}
    \caption{
    Cost of each method, in likelihood evaluations relative to its own cost at the narrowest prior, as the amplitude prior of the AntiGlitch example is widened around a fixed posterior, expressed as the ratio of the prior width to the posterior width; points are the mean of three repeats.
    For Laplace-seeded \aspire, the shading divides the cost between the \ac{DE} \ac{MAP} search (dark) and sampling (light).
    }
    \label{fig:prior-volume}
\end{figure}

\Cref{fig:prior-volume} shows the cost of each method relative to its own cost at the narrowest prior.
The cost of \dynesty grows logarithmically with the prior width, as expected for nested sampling, by 38\% for a prior $10^3$ times wider.
The seeded methods are nearly flat: posterior repartitioning grows by 1.5\% and Laplace-seeded \aspire by 2.4\%.
For Laplace-seeded \aspire, all of this growth is in the \ac{MAP} search, where \ac{DE} must search a wider space (from $7.7\times10^{3}$ to $1.1\times10^{4}$ likelihood evaluations), while the \ac{SMC} stage costs ${\sim}1.2\times10^{5}$ likelihood evaluations at every width.
Prior-initialised \aspire grows fastest, by 63\%, taking one additional tempering iteration (${\sim}3.5\times10^{4}$ likelihood evaluations) for each factor of ${\sim}3$ in the prior width.
For every method, the posterior is unchanged, with a width in $A$ of 0.98--1.03, and the evidence falls by the Occam factor of the prior, with the methods agreeing to within 0.07 when averaged over widths.
These ratios understate the difference in absolute cost: at the narrowest prior, Laplace-seeded \aspire already uses $1.3\times10^{5}$ likelihood evaluations against $1.6\times10^{6}$ for \dynesty, and widening the prior only increases this gap.

The information gain makes the difference between the two \aspire runs explicit: the Kullback--Leibler divergence of the posterior from the reference density $q$ of the tempered target (\cref{eq:tempered}) is 0.1--0.3\,nats at every width for Laplace-seeded \aspire, which reaches $\beta = 1$ in a single step, whereas for prior-initialised \aspire it tracks the divergence of the posterior from the prior, growing from 9 to 16\,nats across the widths, and the number of tempering iterations grows with it at roughly one per nat.
These endpoint divergences ignore the tempering path and the efficiency of the mutations, so they indicate where each run starts rather than the full cost of reaching the posterior.

\subsection{Probability--probability tests}
\label{sec:pp}
To test whether the seeded methods are calibrated across a population, we perform a \ac{PP} test \citep{Cook:2006, Talts:2018, Sidery:2013zua, Veitch:2014wba, Romero-Shaw:2020owr}.
We draw 100 aligned-spin \ac{BNS} signals from the prior and inject each into an independent realisation of Gaussian noise in a network of \ac{LHO} and \ac{LLO} at Advanced LIGO design sensitivity \citep{T0900288} and Virgo and KAGRA at design sensitivity \citep{KAGRA:2020npa}.
We analyse them as for the \ac{BNS} of \cref{sec:simulated}, with \texttt{IMRPhenomD\_NRTidalv2} and a relative-binning likelihood marginalised over distance and phase, but with a distance prior of 10--100\,Mpc, giving a median network \ac{SNR} of 30 (10th--90th percentile 17--49).
We use this network rather than that of \cref{sec:simulated} because there Cosmic Explorer carries almost all of the network \ac{SNR}, so that the sky position is barely constrained by timing between detectors and a failure of sky coverage would say little about the sampler.
Nothing in the analysis is seeded from the injection: the \ac{MAP} search is blind, and the fiducial point of the relative-binning likelihood is found by maximising the likelihood from a draw from the prior.
The secondary-mode search includes the $\psi \to \psi + \pi/2$ symmetry, under which the phase-marginalised likelihood is exactly invariant.
We test Laplace in-prior sampling with both the diagonal and the full truncation, and Laplace-seeded \aspire initialised from the diagonal truncation, with the settings of the \ac{BNS} example (10\,000 particles, 250 \ac{MCMC} steps and up to three modes; \cref{tab:settings}).
Laplace-seeded \aspire completed on 99 of the 100 injections, and we score all three methods on those 99 (see below).
For each injection, we record the credible level at which the injected value lies in each one-dimensional marginal posterior, including the reconstructed distance and phase; for a calibrated method, these are uniformly distributed.

\begin{figure*}
    \centering
    \includegraphics[width=\textwidth]{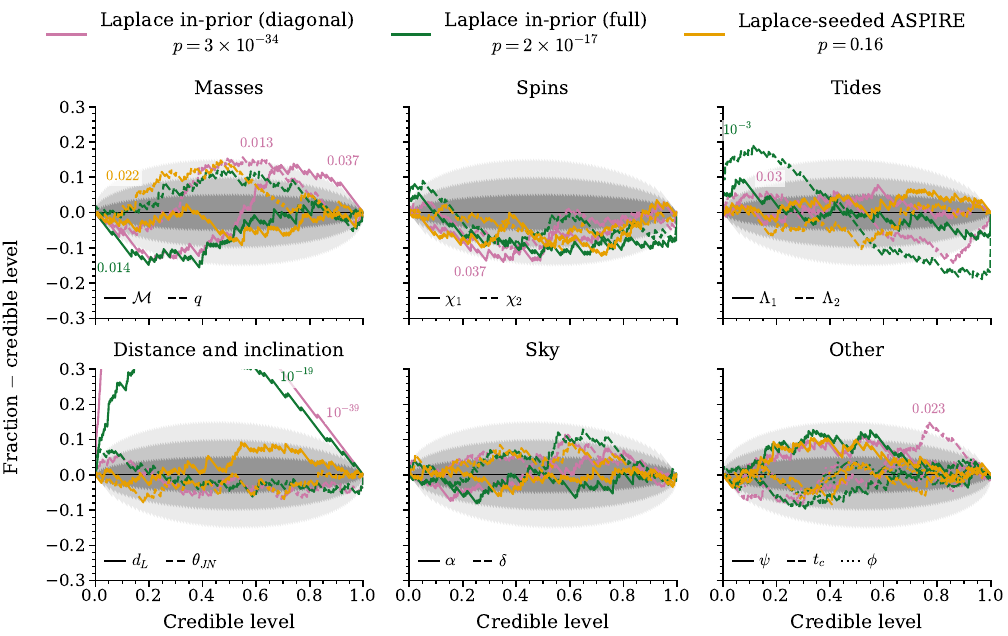}
    \caption{
    \ac{PP} test over 99 simulated aligned-spin \ac{BNS} signals in a four-detector network at design sensitivity, with nothing in the analysis seeded from the injection.
    Each curve is the fraction of injections whose true value lies at or below a given credible level, minus that credible level, so that a calibrated method lies along zero; grey bands show the binomial 1, 2 and 3$\sigma$ spread for 99 injections.
    Colour denotes the method and line style the parameter, grouped into panels; the combined $p$-value of each method is given in the legend, and the Kolmogorov--Smirnov $p$-values of the least consistent parameters are annotated.
    Laplace in-prior sampling fails with either truncation, in opposite directions on different parameters, while Laplace-seeded \aspire passes.
    }
    \label{fig:pp}
\end{figure*}

\Cref{fig:pp} shows the results, with the Kolmogorov--Smirnov $p$-value annotated for the least consistent curves and the combined $p$-value of each method given in the legend.
Laplace-seeded \aspire passes, with a combined $p$-value of 0.16 over 13 parameters; its lowest single $p$-value, 0.022 for the mass ratio, is not unusual for the smallest of 13.
As expected, neither in-prior method passes, with combined $p$-values of $3\times10^{-34}$ (diagonal) and $2\times10^{-17}$ (full): the diagonal truncation discards correlations, so that the chirp mass and merger time are too wide, while the full truncation makes the spins and tidal deformabilities too narrow.
Both fail most strongly in the reconstructed distance, whose true value lies below the 5\% credible level for 49\% (diagonal) and 25\% (full) of the injections.
This is because the in-prior draws fit the data poorly, with the median draw 350 (diagonal) or 230 (full) below the peak log-likelihood, so that the signal appears fainter and the distance larger; restricted to draws that fit the data, the distance credible levels are consistent with uniform.
Laplace-seeded \aspire corrects all of these, at a median cost of $2.2\times10^{7}$ likelihood evaluations (23\,min on 16 cores) per injection, against $2.2\times10^{5}$ (2.0\,min) for in-prior sampling.
On the one injection for which it did not complete, the tempering of \aspire\ stalled after its first step.
A diagnostic rerun suggests that a mutation moved a particle to the boundary of the prior, where the logit transform applied by the flow drives $\ln q$ far below its typical value, so that the particle dominated the incremental weights and no step in $\beta$ larger than the bisection tolerance kept the effective sample size at its target.
This is a robustness limitation of the current implementation, but at least a loud one, not a silently incorrect posterior.

\section{Discussion and conclusion}
\label{sec:discussion}

The properties of gravitational-wave sources observed by current-generation ground-based interferometric detectors are measured using stochastic sampling approaches to approximate the Bayesian posterior distribution.
We investigate the utility of model-informed seeding, based on the Fisher approximation, to accelerate these algorithms.
Our framework uses \ac{DE} to identify the \ac{MAP}, mode searches to identify secondary peaks, and then the \ac{FIM}, within the context of a Laplace approximation, to produce a mixture-model seeding distribution which can be used in two broad ways.

First, samples drawn directly from the seeding distribution (Laplace in-prior sampling) can provide a first-look approximation.
For events observed at \acp{SNR} typical of current-generation observatories, we would not recommend this approach, as the resulting posteriors differ significantly from those obtained by sampling and the event rate is not sufficiently high to justify trading accuracy for speed.
Nevertheless, it could have utility as a tool to help with analysis configuration (e.g. as an initial check of prior bounds and waveform suitability to the approximate posterior).
It may also be useful for glitches, which occur far more often than astrophysical signals and are described by simpler models: for the simulated glitch of \cref{sec:analytic}, Laplace in-prior sampling agrees with \dynesty close to the noise floor at a speed-up of 70--80, which makes it a promising tool for studies of glitch populations \citep{Ashton:2026seh}.
For next-generation networks, at large \acp{SNR}, the Laplace approximation could potentially provide a means to produce rapid posterior estimates.
However, our next-generation examples (\cref{sec:simulated}) show that a louder signal does not by itself make the approximation accurate: the in-prior draws remain far from the posterior in parameters whose posterior is truncated by the prior, such as the tidal deformabilities, and are several times too wide in others, such as $\chi_\mathrm{eff}$ and the sky position of the \ac{BBH}.
These are limitations of a local approximation rather than of its numerical implementation, so producing accurate posterior estimates in this way would require exhaustive identification of secondary modes and a means to capture the non-Gaussian structure that the Laplace approximation misses.

Second, the seeding distribution can be used to seed stochastic sampling algorithms: as a generating distribution for rejection sampling, as an initial distribution for \ac{SMC}, or as a proposal distribution for posterior repartitioning.
We find that for \ac{CBC} events, with our current implementation, rejection sampling is always far too inefficient to be useful.
Seeding \ac{SMC} and posterior repartitioning both produce speed-ups over the current state-of-the-art nested sampling algorithm, \dynesty, but with a trade-off between speed and accuracy.
Laplace-seeded \aspire\ is the faster of the two, with speed-ups of 1.3--3.9 on the compact-binary examples, but it agrees closely with \dynesty\ only for the \ac{BNS} and, in the main peak, GW150914; on the simulated precessing \acp{BBH} it differs by amounts we attribute to low-level sampler systematics, and on GW240615 by considerably more.
Posterior repartitioning agrees with \dynesty\ in all but one example (GW240615), but at more modest speed-ups of 0.6--1.8.
Moreover, prior-initialised \aspire also provides a speed-up: taking that as a baseline, the speed-up from seeding is at most a factor of ${\sim}1.6$, and in four of the six compact-binary examples prior-initialised \aspire has the lower mean intrinsic \ac{JSD}.
Rather, the benefit of seeding over prior-initialised \aspire lies in how the cost scales: prior-initialised \aspire must temper from the prior to the posterior, so its cost grows with the information gain, whereas the seeded run starts close to the posterior and its cost is nearly independent of the prior volume (\cref{sec:prior-volume}), removing the need to restrict the prior to keep the cost down, as is often done for nested sampling.

These speed-ups are obtained with a simple Gaussian-mixture seed and a basic \ac{SMC} configuration, both of which can be improved (see below).
A further reason to consider seeding for future production-style analyses is that it provides a two-stage approach to sampling, in which a rapid approximation of the posterior is followed by a more accurate refinement by stochastic sampling.
This may help in the production of large observational catalogues, in which a first stage provides a quick-look posterior and a second stage a more accurate posterior for events of interest.
The same approach could also be applied to forecasting for next-generation detectors: the \ac{FIM} computed by existing Fisher-based codes \citep[e.g.][]{Iacovelli:2022mbg, Dupletsa:2022scg} could be truncated to the prior support and used to seed \ac{SMC}, producing full posteriors for a population of simulated events, or a representative subset of them.
This would reduce the cost relative to a prior-initialised stochastic sampling of the full population, and would provide a more accurate estimate of the posterior than the Fisher approximation alone.

\emph{Potential improvements}~--- the results presented here depend on three aspects of the method, two of the seeding distribution and one of the configuration of the sampler, and each suggests improvements which may yield better efficiency, better accuracy, or both.
The first is the accuracy of the Fisher approximation within each mode: since \ac{SMC} corrects the seeding distribution exactly only in the limit of infinitely many particles, which we do not reach, a closer match to the posterior reduces both the cost and the residual bias.
With auto-differentiated waveform models and improved numerical methods (for finding the \ac{MAP} and computing the \ac{FIM}), the accuracy of the Laplace approximation could improve, leading to more accurate seeding and potentially larger speed-ups.
Beyond this, the approximation itself could be extended: the \ac{DALI} expansion includes higher-order derivatives of the likelihood to capture non-Gaussian structure that the \ac{FIM} misses \citep{Wang:2022kia}, and a seeding distribution built on such an expansion would match the posterior more closely within each mode, improving both the direct approximation and the performance of the seeded samplers.
The second is the identification of modes: a mode which the search does not find receives little or no weight in the seeding distribution, and the finite sampling stage may not recover it.
For GW150914, Laplace-seeded \aspire under-weights a secondary mode carrying ${\sim}2\%$ of the posterior mass in this way (\cref{sec:events}), a difference which the summary metrics understate and which needs careful consideration in future implementations; until then, seeded runs are not the right tool for current-generation events with significant secondary modes.
Carrying out the search in coordinates in which the posterior is closer to unimodal, such as those of \citet{Roulet:2022kot}, would reduce the number of modes to be found and the non-Gaussian structure within each.
The third is the configuration of \ac{SMC}: we fixed the adaptive schedule of \cref{sec:smc} for all analyses and varied only the numbers of particles and \ac{MCMC} steps (\cref{sec:settings}), but an overly aggressive schedule can lead to weight degeneracy and the loss of modes at a finite number of particles, so a more conservative target efficiency may improve the recovery of minor modes, such as that of GW150914, at the cost of more iterations.

\section*{Data availability}
The data that support the findings of this article are publicly available.
The methods described in this work are implemented in \texttt{bilby\_laplace}~\citep{Ashton:bilbylaplace}, and the scripts used to produce the results, figures and tables are available in the accompanying data release~\citep{Ashton:bilbylaplacedata}.
The strain data for GW150914 and GW240615\_113620 are available from the Gravitational Wave Open Science Center \citep{LIGOScientific:2026jgl}.

\begin{acknowledgments}
The authors would like to thank Colm Talbot for useful discussions in the development of this work and Aditya Vijaykumar for internal LIGO review of the manuscript.
This research has made use of data or software obtained from the Gravitational Wave Open Science Center (gwosc.org), a service of the LIGO Scientific Collaboration, the Virgo Collaboration, and KAGRA.
This material is based upon work supported by NSF's LIGO Laboratory which is a major facility fully funded by the National Science Foundation, as well as the Science and Technology Facilities Council (STFC) of the United Kingdom, the Max-Planck-Society (MPS), and the State of Niedersachsen/Germany for support of the construction of Advanced LIGO and construction and operation of the GEO600 detector.
Additional support for Advanced LIGO was provided by the Australian Research Council.
Virgo is funded, through the European Gravitational Observatory (EGO), by the French Centre National de Recherche Scientifique (CNRS), the Italian Istituto Nazionale di Fisica Nucleare (INFN) and the Dutch Nikhef, with contributions by institutions from Belgium, Germany, Greece, Hungary, Ireland, Japan, Monaco, Poland, Portugal, Spain.
KAGRA is supported by Ministry of Education, Culture, Sports, Science and Technology (MEXT), Japan Society for the Promotion of Science (JSPS) in Japan; National Research Foundation (NRF) and Ministry of Science and ICT (MSIT) in Korea; Academia Sinica (AS) and National Science and Technology Council (NSTC) in Taiwan.
The authors are grateful for computational resources provided by the LIGO Laboratory and supported by National Science Foundation Grants PHY-0757058 and PHY-0823459.
The authors acknowledge the use of the IRIDIS High Performance Computing Facility, and associated support services at the University of Southampton, in the completion of this work.
We acknowledge the use of Claude Opus 5.5 (Anthropic) to assist with code development, result organisation, and manuscript editing.
G.~Ashton is supported by the Science and Technology Facilities Council (STFC) grant UKRI4517.
M.~J.~Williams acknowledges partial support from STFC grant UKRI2490.

\end{acknowledgments}

\appendix

\section{Spikes in the log-posterior}
\label{sec:spikes}
\begin{figure}
    \centering
    \includegraphics[width=\columnwidth]{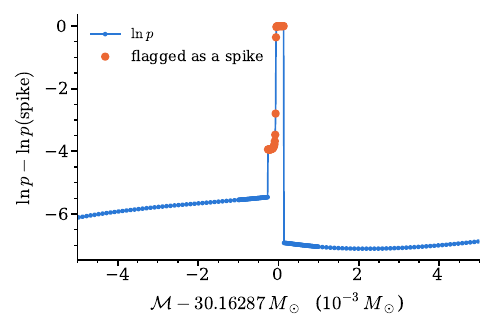}
    \caption{
    The GW150914 log-posterior about a spike found by the \ac{MAP} search, relative to its value at the spike, as a function of chirp mass with all other parameters held fixed.
    Orange points are those flagged by the spike guard of \cref{eq:spike-guard}.
    }
    \label{fig:spikes}
\end{figure}

The \ac{MAP} search discussed in \cref{sec:map} is highly sensitive to structure in the log-posterior that is far narrower than the posterior itself.
Analysing GW150914 \citep{LIGOScientific:2016aoc} with \texttt{IMRPhenomXPHM\_SpinTaylor} \citep{Pratten:2020ceb,Colleoni:2024knd}, we find that about 1\% of single \ac{DE} restarts end on a point standing 1.5--5.5 nats above everything around it.
\Cref{fig:spikes} shows an example, varying the chirp mass $\mathcal{M}$ while holding all other parameters fixed: a staircase about $4\times10^{-4}\,M_{\odot}$ wide, roughly 2000 times narrower than the chirp-mass posterior, beside a step of ${\sim}1.5$ nats.
We first found these on GW150914, where every SpinTaylor version of the precession prescription shows them and the other versions are smooth at the same point, so we take them to be numerical artefacts of that prescription rather than features of the posterior.
They occupy a vanishingly small fraction of the posterior volume, so a stochastic sampler is highly unlikely to visit them, but an optimiser can end on one, and its curvature then describes the spike rather than the posterior.

To mitigate these spikes, we add a \emph{guard} to the \ac{MAP} search.
Working in the unit hypercube on which the prior is uniform, let $e_a(\delta)$ be the amount by which the log-posterior at a \ac{MAP} candidate (the result of one restart) exceeds the larger of its values at offsets $\pm\delta$ along axis $a$; the candidate is flagged as a spike if, along any axis,
\begin{align}
    e_a(\delta) > 1 \quad \text{and} \quad \tfrac{1}{3}\left[4e_a(\delta) - e_a(2\delta)\right] > 1\,,
    \label{eq:spike-guard}
\end{align}
with both thresholds in nats and $\delta = 10^{-3}$.
The second condition extrapolates $e_a$ to zero offset, so that a narrow but smooth peak, for which $e_a(2\delta) \approx 4e_a(\delta)$, cannot trip the test.
Flagged candidates are discarded and the best of the remaining restarts is retained.
On the GW150914 searches above, the guard flags every spike and no other candidate, and we apply it to all \ac{MAP} searches in this work.

\section{Run settings}
\label{sec:run-settings}
\Cref{tab:settings} lists the settings with which each method in \cref{tab:summary} was run.

\begin{table*}
    \centering
    \caption{
    Settings with which each method in \cref{tab:summary} was run, one column per example, grouped into the low-dimensional, \ac{BNS}, simulated \ac{BBH} and observed-event examples.
    A value centred across a group applies to every example in it, and a dash marks a setting that does not apply or a method that was not run.
    Where two \texttt{maxmcmc} values are given, \dynesty\ was run with each.
    The prior-volume study (\cref{sec:prior-volume}) uses a tempering tolerance of $10^{-12}$.
    }
    \footnotesize
    \setlength{\aboverulesep}{0pt}
    \setlength{\belowrulesep}{0pt}
    \renewcommand{\arraystretch}{1.25}
    \resizebox{\textwidth}{!}{
\begin{tabular}{ll|ccc|c|ccc|cc}
\toprule
 & & \multicolumn{3}{c|}{Low-dimensional} & \multicolumn{1}{c|}{BNS} & \multicolumn{3}{c|}{Simulated BBH} & \multicolumn{2}{c}{Events} \\
 & & Gaussian & Rosenbrock & AntiGlitch & BNS 3G & BBH 3G & BBH HLV & BBH2 HLV & GW150914 & GW240615 \\
\midrule
Problem & Sampling dimensions & 2 & 2 & 3 & 11 & \multicolumn{3}{c|}{13} & \multicolumn{2}{c}{13} \\
 & Cores & \multicolumn{3}{c|}{1} & 64 & \multicolumn{3}{c|}{64} & \multicolumn{2}{c}{64} \\
 & Waveform & \multicolumn{3}{c|}{--} & \texttt{IMRPhenomD\_\-NRTidalv2} & \multicolumn{3}{c|}{\texttt{IMRPhenomXPHM}} & \multicolumn{2}{c}{\texttt{IMRPhenomXPHM}} \\
 & $f_\mathrm{ref}$ [Hz] & \multicolumn{3}{c|}{--} & 100 & \multicolumn{3}{c|}{20} & \multicolumn{2}{c}{20} \\
 & Likelihood & \multicolumn{3}{c|}{--} & relative binning & \multicolumn{3}{c|}{standard} & \multicolumn{2}{c}{standard} \\
 & Marginalised & \multicolumn{3}{c|}{--} & $d_L$, $\phi$ & \multicolumn{3}{c|}{$d_L$, $t$} & \multicolumn{2}{c}{$d_L$, $t$} \\
\midrule
Laplace & MAP optimiser & \multicolumn{3}{c|}{DE $\times4$} & DE $\times4$ & \multicolumn{3}{c|}{DE $\times4$} & \multicolumn{2}{c}{DE $\times4$} \\
 & MAP started at injection & \multicolumn{3}{c|}{yes} & yes & \multicolumn{3}{c|}{yes} & \multicolumn{2}{c}{no} \\
 & Curvature & \multicolumn{3}{c|}{Hessian} & Fisher (waveform) & \multicolumn{3}{c|}{Fisher (waveform)} & \multicolumn{2}{c}{Fisher (waveform)} \\
 & Max.\ modes & \multicolumn{3}{c|}{1} & 3 & \multicolumn{3}{c|}{3} & \multicolumn{2}{c}{3} \\
 & Mode searches & \multicolumn{3}{c|}{--} & hypercube, symmetric & \multicolumn{3}{c|}{hypercube, symmetric} & \multicolumn{2}{c}{hypercube, symmetric} \\
 & Search subspace & \multicolumn{3}{c|}{--} & $\kappa$, $\epsilon$ & \multicolumn{3}{c|}{$\kappa$, $\epsilon$, $\delta\phi$} & \multicolumn{2}{c}{$\kappa$, $\epsilon$, $\delta\phi$} \\
 & Symmetries & \multicolumn{3}{c|}{--} & $\psi$ $+\,\pi/2$ & \multicolumn{3}{c|}{$\delta\phi$ $+\,\pi$} & \multicolumn{2}{c}{$\delta\phi$ $+\,\pi$} \\
 & Mode weights & \multicolumn{3}{c|}{--} & laplace & \multicolumn{3}{c|}{laplace} & \multicolumn{2}{c}{laplace} \\
\midrule
In-prior & Draws & \multicolumn{3}{c|}{10\,000} & 10\,000 & \multicolumn{3}{c|}{10\,000} & \multicolumn{2}{c}{10\,000} \\
 & Max.\ modes & \multicolumn{3}{c|}{1} & 3 & \multicolumn{3}{c|}{3} & \multicolumn{2}{c}{3} \\
\midrule
Rejection & Accepted samples & \multicolumn{3}{c|}{10\,000} & -- & \multicolumn{3}{c|}{--} & \multicolumn{2}{c}{--} \\
\midrule
Laplace-seeded & Initial samples drawn from & \multicolumn{3}{c|}{$g(\theta)$} & $g(\theta)$ & \multicolumn{3}{c|}{$g(\theta)$} & \multicolumn{2}{c}{$g(\theta)$} \\
ASPIRE & Seed covariance & \multicolumn{3}{c|}{diagonal} & diagonal & \multicolumn{3}{c|}{diagonal} & \multicolumn{2}{c}{diagonal} \\
  & Initial samples & \multicolumn{3}{c|}{10\,000} & 10\,000 & \multicolumn{3}{c|}{10\,000} & \multicolumn{2}{c}{10\,000} \\
  & Particles & \multicolumn{3}{c|}{5000} & 10\,000 & \multicolumn{3}{c|}{10\,000} & \multicolumn{2}{c}{10\,000} \\
  & MCMC steps / iteration & \multicolumn{3}{c|}{5} & 250 & \multicolumn{3}{c|}{250} & \multicolumn{2}{c}{250} \\
  & Final samples & \multicolumn{3}{c|}{10\,000} & 10\,000 & \multicolumn{3}{c|}{10\,000} & \multicolumn{2}{c}{10\,000} \\
  & Target efficiency & \multicolumn{3}{c|}{0.5--0.8} & 0.5--0.8 & \multicolumn{3}{c|}{0.5--0.8} & \multicolumn{2}{c}{0.5--0.8} \\
  & Tempering tolerance & \multicolumn{3}{c|}{$10^{-6}$} & $10^{-6}$ & \multicolumn{3}{c|}{$10^{-6}$} & \multicolumn{2}{c}{$10^{-6}$} \\
\midrule
Prior-initialised & Initial samples drawn from & \multicolumn{3}{c|}{prior} & prior & \multicolumn{3}{c|}{prior} & \multicolumn{2}{c}{prior} \\
ASPIRE & Initial samples & \multicolumn{3}{c|}{10\,000} & 10\,000 & \multicolumn{3}{c|}{10\,000} & \multicolumn{2}{c}{10\,000} \\
  & Particles & \multicolumn{3}{c|}{5000} & 10\,000 & \multicolumn{3}{c|}{10\,000} & \multicolumn{2}{c}{10\,000} \\
  & MCMC steps / iteration & \multicolumn{3}{c|}{5} & 250 & \multicolumn{3}{c|}{250} & \multicolumn{2}{c}{250} \\
  & Final samples & \multicolumn{3}{c|}{10\,000} & 10\,000 & \multicolumn{3}{c|}{10\,000} & \multicolumn{2}{c}{10\,000} \\
  & Target efficiency & \multicolumn{3}{c|}{0.5--0.8} & 0.5--0.8 & \multicolumn{3}{c|}{0.5--0.8} & \multicolumn{2}{c}{0.5--0.8} \\
  & Tempering tolerance & \multicolumn{3}{c|}{$10^{-6}$} & $10^{-6}$ & \multicolumn{3}{c|}{$10^{-6}$} & \multicolumn{2}{c}{$10^{-6}$} \\
\midrule
Dynesty & Live points & \multicolumn{3}{c|}{1000} & 1000 & \multicolumn{3}{c|}{1000} & \multicolumn{2}{c}{1000} \\
 & Bound & \multicolumn{3}{c|}{\texttt{live}} & \texttt{live} & \multicolumn{3}{c|}{\texttt{live}} & \multicolumn{2}{c}{\texttt{live}} \\
 & Walk & \multicolumn{3}{c|}{acc.-walk, $n_\mathrm{acc}=60$} & acc.-walk, $n_\mathrm{acc}=60$ & \multicolumn{3}{c|}{acc.-walk, $n_\mathrm{acc}=60$} & \multicolumn{2}{c}{acc.-walk, $n_\mathrm{acc}=60$} \\
 & \texttt{maxmcmc} & \multicolumn{3}{c|}{5000, 20\,000} & 5000, 20\,000 & \multicolumn{3}{c|}{5000, 20\,000} & \multicolumn{2}{c}{5000, 20\,000} \\
\midrule
Dynesty + PR & Seed covariance & \multicolumn{3}{c|}{diagonal} & diagonal & \multicolumn{3}{c|}{diagonal} & \multicolumn{2}{c}{diagonal} \\
 & Width inflation & \multicolumn{3}{c|}{2} & 2 & \multicolumn{3}{c|}{2} & \multicolumn{2}{c}{2} \\
\bottomrule
\end{tabular}
}
    \label{tab:settings}
\end{table*}

\section{Walk length of the \dynesty reference}
\label{sec:dynesty-walks}

\begin{figure*}
    \centering
    \includegraphics[width=\textwidth]{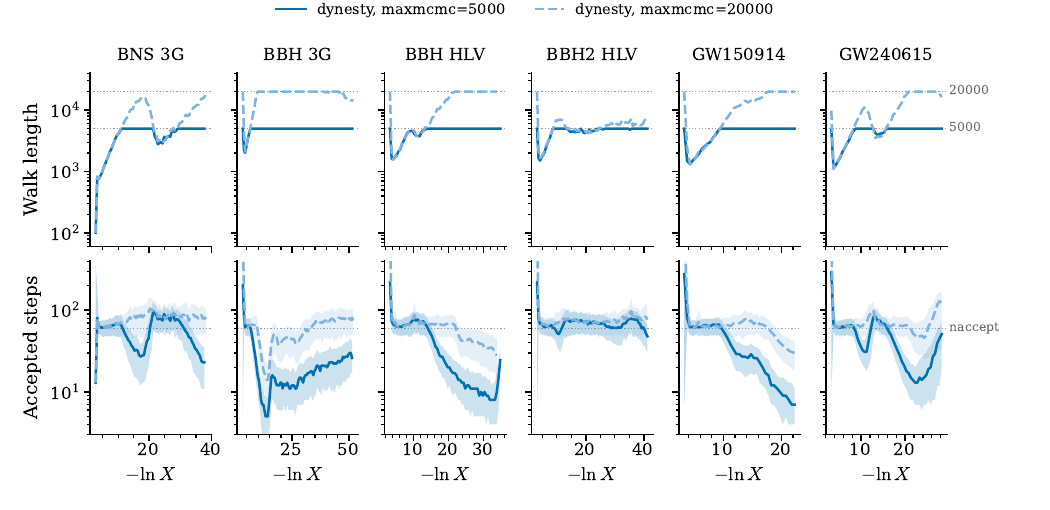}
    \caption{
    Walk length (top) and accepted steps per walk (bottom) of the \dynesty\ reference runs as a function of the compressed prior volume $X$, for the six compact-binary examples of \cref{tab:summary}, at the default cap of \texttt{maxmcmc}${}=5000$ (solid) and at a raised cap of 20\,000 (dashed).
    Lines are medians over iterations in bins of $-\ln X$, and shading the 16th--84th percentile of the accepted steps; dotted lines mark the two caps and the target of 60 accepted steps.
    At the default cap the walks saturate and, except on BBH2 HLV, the accepted steps fall well below the target late in the run; raising it largely restores them, except on BBH 3G, BBH HLV and GW150914.
    }
    \label{fig:naccept}
\end{figure*}

Throughout, the \dynesty\ reference uses the \bilby\ default of \texttt{sample="acceptance-walk"} with a target of 60 accepted steps per walk, and a cap on the walk length of $\texttt{maxmcmc}=5000$.
\bilby\ lengthens each walk until the target is met, but only up to the cap.
Where the cap binds, a new live point is less decorrelated from the one it was copied from than the target intends, which understates the cost of the run and risks an over-tight posterior and a biased evidence.
\Cref{fig:naccept} shows that on the compact-binary examples the cap does bind: at the default, between 54\% and 94\% of iterations are capped, and on BBH HLV, BBH 3G, GW150914 and GW240615 every iteration in the second half of the run is.
The accepted steps then fall well below the target late in the run, to a median of 8--9 over the last 10\% of iterations on BBH HLV and GW150914.
The low-dimensional examples never reach the cap.

We therefore repeated the \dynesty\ runs with the cap raised to 20\,000, with everything else unchanged.
This largely restores the accepted steps, though not on BBH HLV and GW150914 late in the run or on BBH 3G in the middle of it, and costs 1.2--3.6 times as many likelihood evaluations, so that these are the rows with a speed-up below one in \cref{tab:summary}.
On four of the six compact-binary examples, it moves $\ln\evidence$ by at most $1.2\sigma$, and on five it moves the mean \ac{JSD} to the default run by at most 0.25\,mbit above the noise floor.
BBH HLV and BBH 3G are the exceptions in the evidence, which moves by $1.7\sigma$ and $2.1\sigma$, in both cases towards the seeded methods.
Only on BBH HLV does the posterior also move, with a mean \ac{JSD} to the default run of 3.2\,mbit against a noise floor of 2.3\,mbit.
On BBH HLV the seeded methods sit 0.8--1.0 below the default reference but 0.3--0.6 below the raised one, so about half of their evidence offset in \cref{tab:summary} is due to the reference, and on BBH 3G (1.8--2.5 below the default, 1.0--1.7 below the raised) about a third.
The true offset may be smaller still, since BBH HLV and BBH 3G remain under-walked at 20\,000.
We nevertheless keep \texttt{maxmcmc}${}=5000$ as the reference, since this is the standard configuration against which a user would compare, and report the raised-cap runs as separate rows so that both are visible.

\section{Reproducibility and distance to the reference}
\label{sec:bias-variance}

\begin{figure*}
    \centering
    \includegraphics[width=0.49\textwidth]{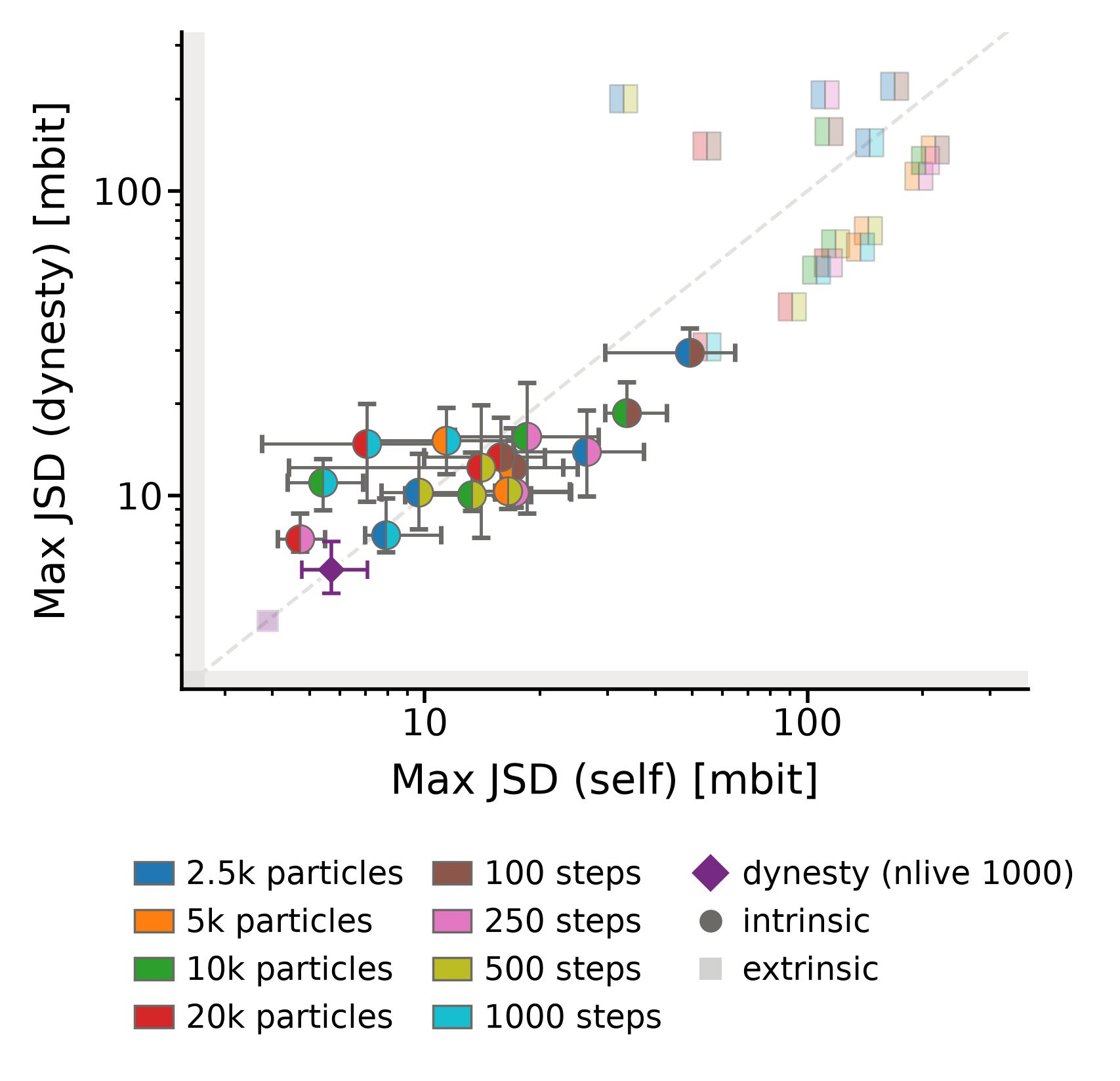}
    \hfill
    \includegraphics[width=0.49\textwidth]{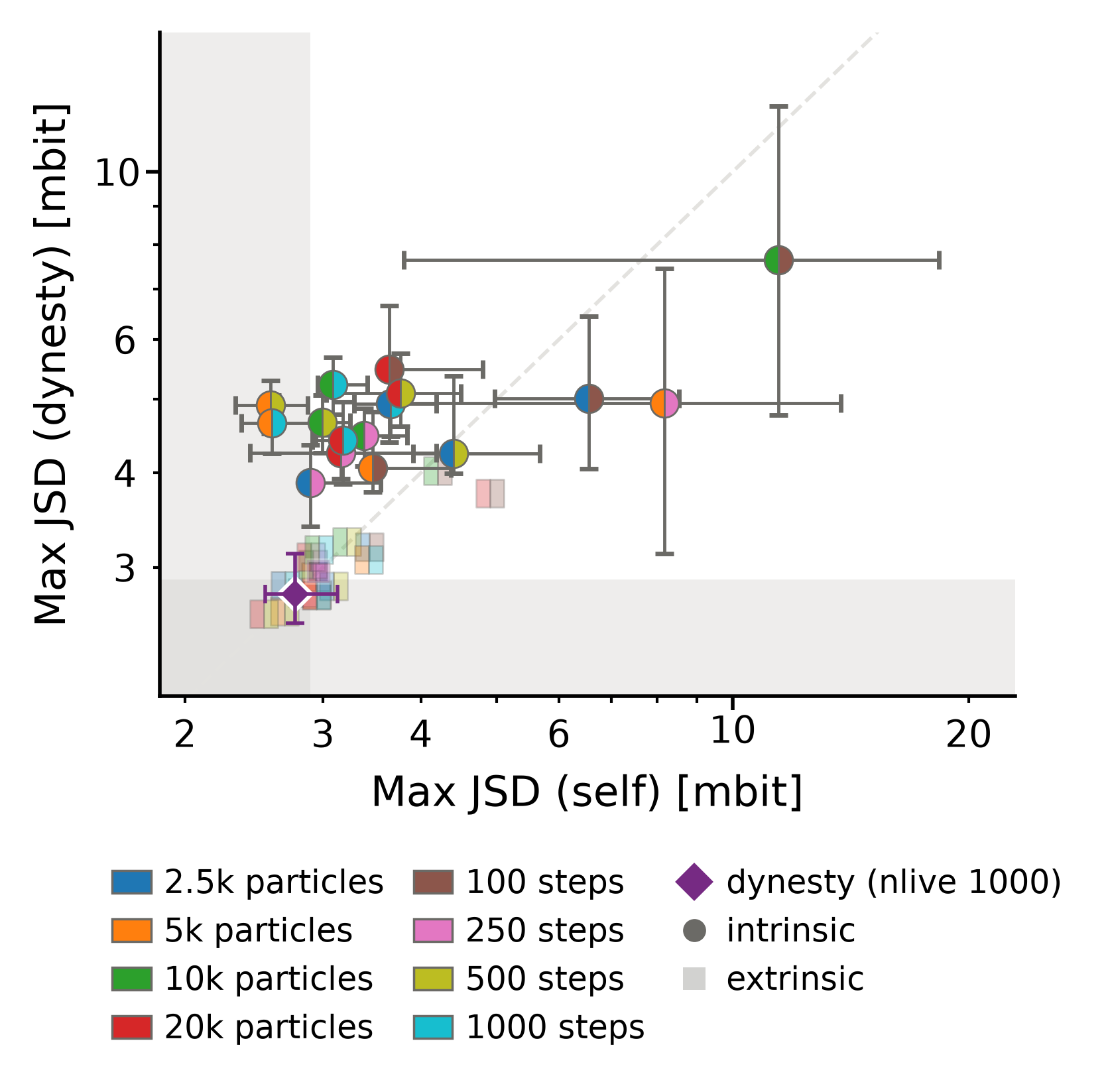}
    \caption{
    Reproducibility against distance to the \dynesty\ reference for the Laplace-seeded \aspire\ configurations of \cref{fig:smc-settings}, for BBH HLV (left) and BNS 3G (right).
    The horizontal axis is the largest one-dimensional \ac{JSD} between the repeats of each configuration, as in \cref{fig:smc-settings}, and the vertical axis the largest one-dimensional \ac{JSD} between its repeats and those of \dynesty\ (1000 live points, \texttt{maxmcmc}${}=5000$).
    Markers and colours are as in \cref{fig:smc-settings}: solid markers take the maximum over the intrinsic parameters, with error bars showing the 16--84\% interval from bootstrap resampling of the repeats, and translucent markers over the extrinsic parameters.
    The grey bands mark the split-half noise floor, and the dashed line equality.
    }
    \label{fig:bias-variance}
\end{figure*}

In \cref{sec:settings}, we chose the \ac{SMC} configuration by its reproducibility, since the true posterior is unknown.
Here we compare that reproducibility with the distance of each configuration from the \dynesty\ reference, which we measure as the largest one-dimensional \ac{JSD} between the repeats of the configuration and those of \dynesty.
\Cref{fig:bias-variance} plots one against the other for the same grid of configurations as \cref{fig:smc-settings}.
A configuration on the diagonal is as far from \dynesty\ as from its own repeats, and one above it differs systematically from \dynesty; \dynesty\ itself lies on the diagonal by construction.

On the intrinsic parameters, the distance to \dynesty\ tracks the reproducibility, up to a small offset.
For BBH HLV, the configurations that reproduce themselves poorly (16--49\,mbit) are also far from \dynesty\ (12--29\,mbit), and no configuration is further from \dynesty\ than from its own repeats by more than ${\sim}8$\,mbit.
For BNS 3G, every configuration lies within 3.9--7.6\,mbit of \dynesty, and those that reproduce themselves as well as \dynesty\ does (2.8\,mbit) still lie 4--5\,mbit from it; for example, 5000 particles with 500 steps reaches 2.6\,mbit against its own repeats but 4.9\,mbit against \dynesty.
On these parameters, then, a configuration that reproduces itself is also close to the reference, and reproducibility is a useful, though not complete, test of a run.

In each case, the offset is concentrated in a single parameter: the spin magnitude $a_1$ for BBH HLV (${\sim}9.5$\,mbit to \dynesty\ against ${\sim}3.5$\,mbit between \ac{SMC} repeats) and $\chi_1$ for BNS 3G (4.2--5.2\,mbit against 2.1--3.0\,mbit).
For BNS 3G, the \ac{SMC} posteriors are wider in $\chi_1$ than those of \dynesty, with a standard deviation of 0.0121--0.0123 against 0.0113, and a single \dynesty\ run with the walk-length cap raised to 20\,000 (\cref{sec:dynesty-walks}) lies between the two, at 0.0118, while the \dynesty\ run under the repartitioned prior (\cref{sec:pr}) matches \ac{SMC}, at 0.0121.
This suggests that the \dynesty\ reference at the default cap is slightly too narrow in $\chi_1$, rather than that \ac{SMC} is biased, though a single run does not settle this.

Reproducibility does not detect an error that every repeat shares.
On the extrinsic parameters of BBH HLV, 2500 particles with 500 steps reproduces itself at 33\,mbit but lies 200\,mbit from \dynesty, because all five of its repeats drop the $\delta\phi$ lobe, and so agree with each other.
A multimodal posterior that the sampler cannot traverse is therefore the case in which reproducibility is not enough, and we still compare against an independent reference wherever one is available.

\section{Mass-ratio ridge of GW150914}
\label{sec:gw150914-ridge}
\begin{figure}
    \centering
    \includegraphics[width=\columnwidth]{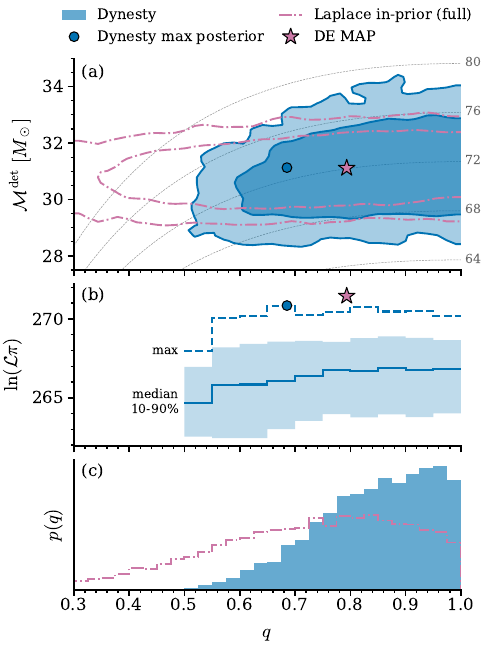}
    \caption{
    As \cref{fig:gw240615-ridge}, for GW150914, with curves of constant total mass every $4\,M_\odot$.
    }
    \label{fig:gw150914-ridge}
\end{figure}

\Cref{fig:gw150914-ridge} repeats \cref{fig:gw240615-ridge} for GW150914.
The total mass is measured and the likelihood is flat along the line of constant total mass, as on GW240615, but the ridge stops at $q \approx 0.6$, and the \ac{MAP} and the highest-posterior \dynesty\ sample both lie on its flat top.
Below it the likelihood falls off, so that $q$ is better measured than on GW240615 (posterior standard deviation 0.11 against 0.13), and at the \ac{MAP} the curvature along $q$ is below that of the prior, so the width of the Gaussian is capped at the prior's.
The in-prior draw then covers too much of $q$, not too little: it spans $0.40$--$0.97$ against $0.64$--$0.98$ for \dynesty, with 88\% of the \dynesty\ samples inside its 5--95\% range, against 13\% on GW240615.
This costs the in-prior draw some \ac{JSD} (139--146\,mbit in $q$), but it is the safe direction for \ac{SMC} and posterior repartitioning, which only need to contract it.
The evidence deficit of the in-prior draw, by 4--10 on every gravitational-wave example in \cref{tab:summary}, does not distinguish the two events, and is not a signature of the ridge.

\bibliography{bibliography}

\end{document}